\documentclass{article}

\usepackage{graphicx,amssymb,amsmath,amsthm}

\usepackage{booktabs} 
\usepackage{array}    
\usepackage{xcolor}
\usepackage{lineno} 
\usepackage{soul} 
\usepackage[normalem]{ulem}
\usepackage[margin=1in]{geometry}

\usepackage{enumitem}
\usepackage{hyperref}
\usepackage{algorithm}
\usepackage{algorithmic}
\usepackage{cancel}
\usepackage{comment}

\newtheorem{theorem}{Theorem}
\newtheorem{lemma}[theorem]{Lemma}
\newtheorem{corollary}[theorem]{Corollary}

\newtheorem{proposition}[theorem]{Proposition}

\title{Catenary Visibility: A Geometric Framework\\ for Planning with Tethered Robots}

\author{
B. Ábrego\thanks{Department of Mathematics, California State University, Northridge, {\tt\small [bernardo.abrego,silvia.fernandez]@csun.edu}} , 
J.M. D\'iaz-B\'añez\thanks{Department of Applied Mathematics II, University of Seville, Spain, {\tt\small [dbanez,vscanales]@us.es}} ,
S. Fernández$^*$,
D. Flores-Pe\~naloza\thanks{Facultad de Ciencias, Universidad Nacional Aut\'onoma de M\'exico, {\tt\small dflorespenaloza@ciencias.unam.mx}}, 
V. Sánchez-Canales$^\dagger$
}

\date{}

\begin{document}

\maketitle

\begin{abstract}
This paper introduces \emph{catenary visibility}, a new geometric visibility concept for a robotic system consisting of a ground robot tethered to an aerial drone. The tether is modeled as an extensible catenary curve with bounded length, providing sustained power and significantly extending operational range. We define a point on the ground plane as \emph{catenary-visible} from a target point in 3D if there exists a collision-free catenary connecting them, given a set of obstacles. This paper introduces a geometric framework for visibility-aware planning of a tethered marsupial system, addressing the computational challenge of determining the feasible deployment locus: the collection of points in the ground plane that are catenary-visible from a given target in 3D space. We begin by solving the 2D subproblem, where the obstacles, target, and the ground vehicle's line of motion are all constrained to a single vertical plane. Our algorithm for this case computes the complete set of feasible launch points along the motion line and simultaneously determines the range of viable cable lengths for each. We also consider the query problem of finding all collision-free length ranges for a given location of the ground robot. Building upon this planar solution, we then generalize the approach to three dimensions, deriving a 3D solution for a constrained class of obstacles.
\end{abstract}

\section{Introduction}

The profound impact of drone technology on modern society is undeniable. Yet, a critical bottleneck persists: the limited endurance of small aerial platforms restricts their use in long-term operations. Marsupial robotics—where the aerial vehicle is deployed from and sustained by a ground vehicle—emerges as a strategic paradigm to overcome this limitation, enabling persistent aerial presence for surveillance, inspection, rescue, or delivery tasks.
A \emph{tethered marsupial robotics system} consists of three components: an Unmanned Ground Vehicle (UGV), an Unmanned Aerial Vehicle (UAV), and a tether that connects them, supplying energy to the UAV. These systems are particularly advantageous in industrial applications, as they significantly extend the UAV’s battery life during flight.
Computing collision-free trajectories for a marsupial system is a challenging problem in robotics, specifically when the catenary curve is used to model the shape of a hanging tether \cite{martinez-rozas_icra21, martinez2023path,battocletti2024entanglement}.  The catenary model is inherently complex, and answering collision avoidance queries during the planning process can be time-consuming. Therefore, determining the locus of feasible take-off points for the UAV becomes a critical problem. This fact becomes crucial for path planning with a marsupial system in a decoupled approach where the UGV first
moves to a position from which the UAV can take-off and fly through a vertical plane to
reach an aerial target. Identifying these points in advance allows us to determine the optimal UGV stopping locations for UAV launch and avoids solving the query problem on collisions between the catenary curve and the obstacles, allowing more efficient path planning procedures in complex scenarios \cite{martinez2025long}. 

In the 2D setting, our problem reduces to the following task: given a vertical plane with obstacles, a target point $A$, and a horizontal line $\cal L$, find all points on $\cal L$ from which a catenary of maximum length $L$ can reach $A$ without intersecting any obstacle. This question provides us with a new concept of geometric visibility, which we will call \emph{catenary visibility}, and the question is: what points of $\cal L$  \emph{are catenary-visible} from the target $A$? The problem is formulated exactly the same way in a three-dimensional scenario where the UGV moves on a horizontal plane. The \textbf{Catenary Visibility Problem} was recently identified as an open challenge in \cite{capitan2026maspa}. In that work, the authors state that ``an exact algorithm for this problem could reduce execution time and improve path lengths."

The problem of catenary visibility is more complex than that of direct line-of-sight. Under direct visibility, the region occluded by a single 3D obstacle is simply a shadow—a cone-like projection. In contrast, catenary visibility is constrained by feasible catenary curves, meaning the non-visible region on the ground plane can be non-convex and geometrically intricate. Moreover, catenary visibility provides flexibility as there are families of obstacles with no direct visibility that actually have (partial o total) catenary visibility. In fact the direct visibility solution is always contained in the catenary visibility solution.

\begin{figure}
    \centering
    \includegraphics[width=0.4\linewidth]{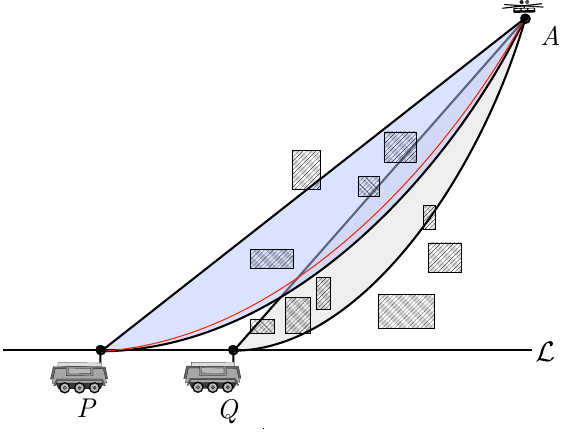}
    \caption{In this situation, point $P$ is catenary-visible from $A$, whereas point $Q$ is not.}
    \label{fig:car}
\end{figure}

A common modeling simplification in robotics is to represent obstacles as axis-aligned bounding volumes, specifically, cuboids in 3D or rectangles in 2D \cite{khan2014rmap}. These shapes are typically chosen as the minimal bounding boxes that completely contain the corresponding obstacle.
In this work, we assume orthogonal rectangles in 2D and boxes in 3D in the shape of specific wedges that allow the extension of the procedure developed in 2D. With this assumption, we propose a method that not only obtains the geometric location of the catenary-visible points, but can also respond to the range of catenary lengths that make it possible. Figure \ref{fig:car} depicts a two-dimensional scenario involving two positions, $P$ and $Q$. While a range of feasible catenary lengths exists for position $P$, point $Q$ remains inaccessible under the given catenary visibility constraints, regardless of the cable length considered.

In the following sections, we summarize related work, formally define the problem, and solve the 2D version. We then show how to extend the approach to 3D. Finally, we discuss extensions and open problems in the last section. For easy reference, a list of notation is included in the Appendix.

\section{Positioning Our Approach Within Related Work}\label{sec:RW}

Marsupial robotic systems —comprising an Unmanned Ground Vehicle (UGV) and an Unmanned Aerial Vehicle (UAV) connected by a power- supplying cable— significantly enhances the UAV's operational efficiency while enabling access to hard-to-reach inspection sites, eliminating the need for frequent battery replacements. This design presents a complex, hierarchical path planning problem that sits at the intersection of robotic coordination and computational geometry. The carrier’s path must not only be efficient and feasible but must also serve as a dynamic platform for maximizing the coverage and operational effectiveness of its cubs. This inherently geometric problem demands reasoning about visibility, occlusion, and region decomposition in both 2D and 3D spaces. Although the specific concept of \emph{marsupial visibility} is novel, it comes from foundational work in tether-aware planning,  geometric motion planning, and visibility and shadow computation.
This review synthesizes relevant work from robotic path planning and geometric algorithms to establish the foundation and identify gaps in marsupial-specific planning.

The primary motivation for tethering a UAV to a ground vehicle is to overcome the severe endurance limitations of battery-powered flight. The core planning challenge for marsupial teams is deciding the route of the UGV, the deployment/retrieval points, and the detailed task paths of the drone. The literature reveals a spectrum of planning architectures, evolving from simple decoupled methods \cite{grocholsky2006cooperative} to more sophisticated integrated models \cite{martinez2023path}.     Decoupled planning methods offer greater efficiency for missions focused on reaching a specific target or set of targets, whereas coupled planning frameworks are better suited for exploration and coverage tasks where the objective is to maximize information gain or map an unknown region. 
In these scenarios, the tether’s physical properties must be incorporated into the planning model. It can be abstracted as a tight cable \cite{xiao2021autonomous,lima2025tension} for computational simplicity, but a more realistic representation is the catenary curve \cite{martinez-rozas_icra21}, which accounts for sag due to gravity and dynamic swing. This higher-fidelity model is essential for accurate collision checking and the generation of kinematically feasible paths. A comprehensive framework for integrating such a tether model directly into the motion planning stage is presented in \cite{battocletti2024entanglement}.

In the classical decoupled planning paradigm, the carrier's path to a set of predefined deployment points is planned first, typically using graph-based algorithms such as Dijkstra or $A^*$ over a discrete representation of the environment (see, for example, \cite{kim2025hierarchical}). Subsequently, a separate, local planning problem is solved for the deployed agent (e.g., a micro-air vehicle or drone) operating within a constrained radius from its drop-off point. This approach, however, introduces a critical and computationally expensive geometric verification step: at each candidate deployment point, one must determine whether a collision-free catenary exists to reach the intended target. In such a case, the drone's trajectory could follow the catenary curve by integrating the planner
with controllers for flight stabilization and trajectory tracking \cite{xu2023event}.
Representing the tether as a limited length catenary curve—rather than a simple straight line—requires solving a collision free catenary problem at each point to check for ground or obstacle interference. Performing this analysis online, for every candidate point during path search, is prohibitively time-consuming and negates the efficiency benefits of decoupled planning.

To overcome this bottleneck, our work proposes a geometric preprocessing stage. We compute the feasible deployment space that is the 2D region of carrier poses from which a tether-based deployment to a given target region is  collision-free. The result is a precomputed, query-able map that the high-level planner can use to instantly validate deployment points. This transforms an expensive, continuous collision-checking problem into a fast table lookup, thereby restoring planning efficiency while maintaining rigorous physical fidelity.

Prior to our work, a recent approach 
\cite{capitan2026maspa} discretizes the state space to approximate an optimal path that minimizes the combined length of the ground vehicle and UAV trajectories. A key contribution is the efficient computation of a feasible take-off region for the UAV when the tether is modeled as a taut polygonal curve.  This solution is then applied as a constraint to identify a discrete set of feasible points under a more realistic catenary tether model. Their resulting planner, the Multi-Agent Shortest Path Algorithm (MASPA), demonstrates strong performance in both random and realistic scenarios. Notably, it outperforms a state-of-the-art RRT*-based baseline in terms of total path length and computational execution time.
In that paper, the authors conclude that computing the exact locus of visible points with a catenary curve
is a challenging geometrical problem.  An exact algorithm for this problem could reduce the execution time and
improve the path lengths. 
Our approach responds to that open question of computing the exact locus of visible points
in the 2D and 3D space, being a crucial tool for designing an efficient and effective online path planning algorithm.

Finally, we mention that the problem addressed in this paper computes shadow regions to automatically identify optimal deployment points for the tethered agent and can be related to the Art Gallery Problem (AGP).
The computation of visibility (what can be seen from a given point) and shadows (the occluded complement) is a foundational problem at the intersection of computational geometry \cite{ghosh2007visibility,de2008computational}, computer graphics \cite{bittner2003visibility}, and robotics \cite{wang2015visibility}. 
In fact, the decision of where to deploy a cub is fundamentally a geometric problem of maximizing visibility from a set of aerial guarded locations
\cite{gartner2025optimizing}.
The classic AGP, which seeks the minimum number of guards (sensors) to see every point in a polygon \cite{o1987art}, has inspired numerous robotic coverage algorithms. For mobile sensors, the problem becomes the Mobile Guard or Watchman Route Problem, where the goal is to find the shortest path for a moving guard to see the entire environment \cite{chin1991shortest,tan2004approximation}. In marsupial systems, the carrier could act as a mobile guard that can place static guards (the cubs). Our problem can be considered, in a certain sense, one of inverse visibility, that is, we need to see a point or a set of points in the space, and we seek to calculate the points on the plane from which the given target points can be ``seen'' with a catenary curve. We believe that this could be the source of new visibility problems in two and three dimensions. 

\section{The 2-dimensional scenario}
\label{sec:2D}

\subsection{Problem statement}

 Let $L>0$ and $\pi$ be a vertical plane containing a set of rectangular obstacles ${\cal R}=\{R_1,R_2,\dots,R_n\}$. Additionally, let $\cal L$ and $A$ denote a line and a point (referred to as \emph{target point}) in $\pi$, respectively.
A curve ${\cal C}$ in $\pi$ is collision-free if ${\cal C}$ does not intersect of any obstacle in ${\cal R}$. We say that a point $G$ in $\cal L$ is \emph{catenary-visible}, if there
exists a collision-free catenary connecting $G$ and $A$ with
length less than or equal to $L$. 
We define the \emph {2-dimensional
catenary visibility problem} (2D-CV problem) as the problem of computing the locus of the catenary-visible points in $\cal L$. 

In this section, we present a method for solving the 2D-CV Problem, which also calculates all the length values of a collision-free catenary for each catenary-visible point in $\cal L$.

\subsection{Our model}

\begin{figure}[htb]
\centering
    \begin{minipage}[b]{0.47\textwidth}
    \centering
    \includegraphics[width=1\linewidth]{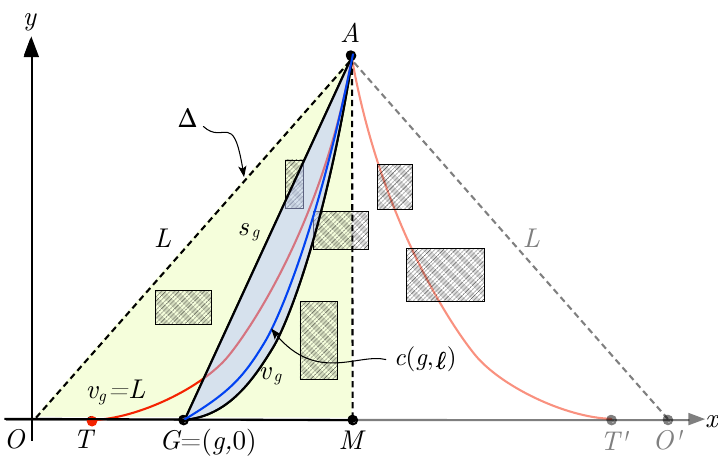}
    \caption{The obstacle plane. $G$ represents the current position of the UGV and $A$ is the target point of the UAV. By symmetry, we only consider obstacles in the shaded closed region $\Delta$. $T$ is the transition point,  the (blue) shaded region corresponds to the possible extensions of the tether (catenaries) from $G=(g,0)$ to $A$, $s_g$ and $v_g$ are, respectively, the shortest and longest lengths of all these extensions.}
    \label{fig:model1}
\end{minipage}
\hfill
\begin{minipage}[b]{0.47\textwidth}
    \centering
    \includegraphics[width=.85\linewidth]{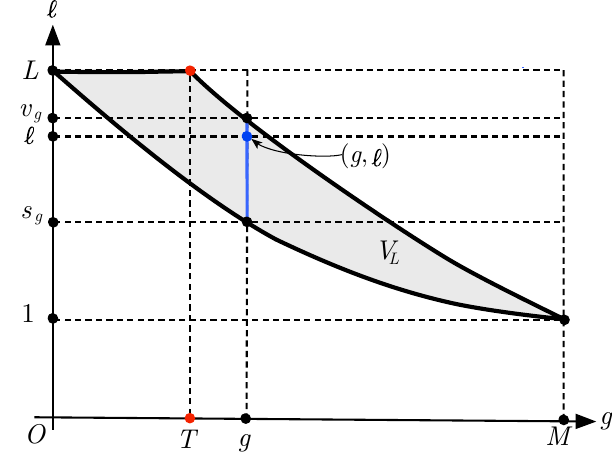}
    \vspace{.14in}
    \caption{The (catenary) visibility plane. The visibility region $V_L$ does not take into account any obstacles. The point $(g,\ell)$ corresponds to the catenary $c(g,\ell)$ in the obstacle plane. The blue vertical segment through $(g,\ell)$ corresponds to all possible lengths of the tether when the UGV is at position $G=(g,0)$ in the obstacle plane. It is bounded by the points $(g,s_g)$ below and $(g,v_g)$ above.}
    \label{fig:visibility1}
\end{minipage}
\end{figure}

In this model, we assume that the UGV moves in the $xy$-plane (the obstacle plane $\pi$) along the positive $x$-axis (the line $\cal L$), the tether has maximum length $L$, and the UAV's location is represented by the target point $A=(\sqrt{L^2-1},1)$, see Figure \ref{fig:model1} . This means that the tether is completely extended when the UGV is at the origin $O=(0,0)$ or at the point $O'=(2\sqrt{L^2-1},0)$. Let $M=(\sqrt{L^2-1},0)$ be the midpoint of $OO'$, i.e., the foot of the perpendicular axis from $A$ to the $x$-axis.

We use the point $G=(g,0)$ to represent the location of the UGV, where $0\leq g\leq 2\sqrt{L^2-1}$.  At any moment, the tether naturally describes a catenary curve or, when tight, a straight line segment. We think of these straight line segments as \emph{degenerate catenaries} and often refer to them simply as catenaries to avoid separately considering them. When the UGV is at position $G=(g,0)$ and the tether is extended to length $\ell$, we use $c(g,\ell)$ to denote the catenary described by the tether from $G$ to $A$. In our model, the drone flies through the curve defined by the catenary, from point $G$ to point $A$, while the UGV continuously supplies tether so that it also coincides with the trajectory marked by the catenary. 

Let $s_g$ be the length of the segment $GA$ and $v_g$ the length from $G$ to $A$ of the catenary with vertex $G$ passing through $A$. We ask for the length of the tether $1\leq \ell\leq L$ to be such that the portion of the catenary between $G$ and $A$ does not go below the $x$-axis. This means that for certain points in $\cal L$, namely when $v_g<L$, the tether cannot be extended to its maximum length $L$ but only to length $v_g$. When $G=O$ or $O'$ the tether is forced to be a straight line segment of length $L=s_0=s_{2\sqrt{L^2-1}}$; and when $G=M$, the tether is forced to be a straight line segment of length $1=s_{\sqrt{L^2-1}}$. For all other positions, the length of the tether can go from  $s_g$ to $\min(v_g,L)$. For a given feasible $L$, let $T=(t,0)$ be the \emph{transition point} in $OM$ such that $v_t=L$.  The following propositions shows that this transition point is unique, and gives its explicit value. 

\begin{proposition}
Let $C$ be a catenary whose vertex is some point $T \in OM$ passing through point $A$, as defined previosly. Let $L$ be the lenght of the arc of this catenary between $T$ and $A$. Then, the coordinates of $T$ are
\[
T=\left(\sqrt{L^2-1}-\frac{{L^2-1}}{2}\ln\left(\frac{L+1}{L-1}\right),0\right).
\]
\end{proposition}
\begin{proof} 
The catenary $C$ has the equation
\begin{equation}\label{eq:catenary}
    y+a=a\cosh\left(\frac{x-t}{a}\right), 
\end{equation}
for some $a>0$. The length $s$ of the arc of $C$ from a point $(x_1,y_1)$ to its vertex $(x_0,y_0)$, satisfies 
\begin{equation}\label{eq:distance}
    y_1-y_0+a=\sqrt{a^2+s^2}.
\end{equation}
 
It follows that 
    $a=\frac{L^2-1}{2}$.
Since $A$ is a point on this catenary, we have 
\[
\frac{1+a}{a} =\cosh\left(
\frac{\sqrt{L^2-1}-t}{a} 
\right) .
\]
The result follows by standard algebra.

\end{proof}

Any object $R$ that intersects the triangle $OAO'$ is considered an \emph{obstacle}, since in this case there exists a catenary $c(g,\ell)$ that intersects $R$.  In this case, we say that $R$ is an \emph{obstruction} for $c(g,\ell)$ or that $c(g,\ell)$ is \emph{obstructed} by $R$. By symmetry about the line $AM$, we restrict our attention to obstacles contained in the triangle $\Delta=\triangle(OAM)$. Whenever the catenary $c(g,\ell)$ is not obstructed by any obstacle, we say that the point $G=(g,0)$ is \emph{catenary-visible}, or simply \emph{visible} from $A$ (through $c(g,\ell)$). The \emph{visibility region} for a set of obstacles is the set of catenaries $c(g,\ell)$ in the obstacle plane, or equivalently, the set of points $(g,\ell)$ in the visibility plane, with $0\leq g\leq \sqrt{L^2-1}$ for which $G$ is visible, see Figure \ref{fig:visibility1}. We are interested in determining this visibility region for a given finite set of obstacles contained in $\Delta$. In the special case where there are no obstacles present, we denote this visibility region by $V_L$. Whenever there is an obstacle $R$ or a family  of obstacles ${\cal R}$ present in $\Delta$, the subset of $V_L$ obstructed by $R$ or ${\cal R}$ is denoted by $R^-$ and ${\cal R}^-$, respectively; and thus the corresponding visibility regions are $V_L\setminus R^-$ and $V_L\setminus {\cal R}^-$. In fact, to complete the preprocessing geometric stage (determining the feasible deployment space), we only need to determine the set of values of $g \in [0,\sqrt{L^2-1}]$ for which there is a \emph{feasible length} $\ell\in [s_g,min(v_g,L)]$ such that there is visibility through the catenary $c(g,\ell)$. The set of all feasible lengths for $g$ is denoted by $F_g$. We say that $g$ has \emph{null visibility} if $F_g=\emptyset$, \emph{full visibility} if $F_g=[s_g,min(v_g,L)]$, and \emph{partial visibility}, otherwise. The \emph{shadow} of a set of obstacles ${\cal R}$ is the set $\{g\in [0,\sqrt{L^2-1}]:F_g=\emptyset\}$ and its complement is called the \emph{visibility range} of $\cal R$.

\subsection{The visibility region with no obstacles present}
In this section, we explicitly describe the visibility region $V_L$ when no obstacles are present in $\Delta$ as shown in Figure \ref{fig:visibility1}.  We start by proving a lemma  about catenaries.

\begin{lemma}\label{lem:unique_catenary_3}
    Let $P, Q$, and $Z$ be three points in the plane. Then there is at most one catenary of the form $y-k=a(\cosh{\frac{x-h}{a}}-1)$ passing through $P$, $Q$ and $Z$. 
\end{lemma}
\begin{proof}
    Suppose that $y-k_1=a_1(\cosh{\frac{x-h_1}{a_1}}-1)$ and $y-k_2=a_2(\cosh{\frac{x-h_2}{a_2}}-1)$ are catenaries through $P$, $Q$ and $Z$. Let $f$ and $g$ be isometries that take these two catenaries to the standard catenary $y=\cosh{x}$, respectively. These isometries exist because any two catenaries of such a form are similar. Each of them is in fact a dilation followed by a translation. Then $\triangle (PQZ)$, $\triangle (f(P)f(Q)f(Z))$, and $\triangle (g(P)g(Q)g(Z))$ are directly similar and their corresponding sides are parallel. We prove that the last two triangles are indeed the same, which in turn proves that $f=g$ and so the two catenaries are equal. Since pairs of corresponding sides of $\triangle (f(P)f(Q)f(Z))$ and $\triangle (g(P)g(Q)g(Z))$ are parallel, they do not intersect. In particular (since all the vertices are in the same catenary), their projections to the $x$-axis are either one contained in the other or completely disjoint. Since all three pairs cannot satisfy that their projections are not one contained in the other, then there is at least one pair whose projections are disjoint. Suppose without loss of generality that such pair of segments is $f(P)f(Q)$ and $g(P)g(Q)$ and that their $x$-coordinates satisfy $f(P)_x<f(Q)_x< g(P)_x<g(Q)_x$. By the Mean Value Theorem applied to $y=\cosh{x}$, there are $f(P)_x< c_1<f(Q)_x$ and $g(P)_x<c_2<g(Q)_x$ such that $y'(c_1)=\sinh{c_1}$ is the slope of the line $f(P)f(Q)$ and $y'(c_2)=\sinh{c_2}$ is the slope of $g(P)g(Q)$. Since these two lines are parallel and $\sinh{x}$ is strictly increasing, then $\sinh{c_1}=\sinh{c_2}$ and so $c_1=c_2$ contradicting that the intervals $[f(P)_x,f(Q)_x]$ and $[g(P)_x,g(Q)_x]$ are disjoint. 
\end{proof}
The following result describes the visibility region $V_L$.

\begin{proposition}\label{prop:V_L}
    The visibility region $V_L$, when no obstacles are present in $\Delta$, is the closed region bounded below by the set $\beta_1$ and above by the set $\beta_2\cup\beta_3$, where 
    \begin{align*}
        \beta_1 & =\left\{(g,s_g): 0\leq g \leq \sqrt{L^2-1}\right\}\\
        \beta_2&=\left\{(g,L): 0\leq g \leq t\right\},\\
        \beta_3&=\left\{(g,v_g): t\leq g \leq \sqrt{L^2-1}\right\},\\
        s_g&=\sqrt{L^2+g^2-2g\sqrt{L^2-1}},\\
        t&=\sqrt{L^2-1}-\frac{{L^2-1}}{2}\ln\left(\frac{L+1}{L-1}\right),
    \end{align*}
    and $v_g$ is given by the implicit equation 
    \begin{align*}
     g=\sqrt{L^2-1}-\frac{v_g^2-1}{2}\ln{\left(\frac{v_g+1}{v_g-1}\right)}.
\end{align*}
\end{proposition}

\begin{proof}
    The region $V_L$ consists of all points $(g,\ell)$ such that $0\leq g\leq \sqrt{L^2-1}$ and $s_g\leq \ell\leq \min(v_g,L)$, that is, $\ell$ is a possible length of the catenary connecting $G=(g,0)$ and $A=(\sqrt{L^2-1},1)$. Hence, this region is bounded below by the curve corresponding to the length $s_g$ of the segment $GA$, that is, 
    \begin{equation*}
        \ell=s_g=\sqrt{1+(\sqrt{L^2-1}-g)^2}=\sqrt{L^2+g^2-2g\sqrt{L^2-1}}.
    \end{equation*}
    The region $V_L$ is bounded above by the horizontal line $\ell=L$ when $0\leq g \leq t$ and by the curve $\ell=v_g$ when $t\leq g \leq \sqrt{L^2-1}$. In this case, the length $v_g$ decreases continuously from $L$ to $1$ and is given implicitly by 
\begin{align*}
     g=\sqrt{L^2-1}-\frac{v_g^2-1}{2}\ln{\left(\frac{v_g+1}{v_g-1}\right)},
\end{align*}
which is obtained from the equation of the catenary with vertex $(g,0)$ applied to point $A=(\sqrt{L^2-1},1)$ (see Equation (\ref{eq:catenary}))
\begin{equation}\label{eq:catenary2}
    1+a=a\cosh\left(\frac{\sqrt{L^2-1}-g}{a}\right)
\end{equation}
and the equation  
\begin{equation}\label{eq:distance2}
1+a=\sqrt{a^2+v_g^2}
\end{equation}
satisfied by the length $v_g$ of such catenary between its vertex $G$ and the point $A$ (see Equation (\ref{eq:distance})). From Equation (\ref{eq:distance2}), we get $a=(v_g^2-1)/2$, which together with (\ref{eq:catenary2}) gives
$$1+\frac{v_g^2-1}{2}=\frac{v_g^2-1}{2}\cosh\left(\frac{2\sqrt{L^2-1}-2g}{v_g^2-1}\right)$$
and then, 
$$g=\sqrt{L^2-1}-\frac{v_g^2-1}{2}\ln{\left(\frac{v_g+1}{v_g-1}\right)}.$$

We now argue that the area between $c(g,s_g)$ and $c(g,\min (L,v_g))$  is continuously swept as the catenaries through $g$ increase in length, which means that the entire vertical segment from $(g,s_g)$ to $(g,\min(L,v_g))$ is contained in $V_L$. For each $0\leq g\leq \sqrt{L^2-1}$, let $C_g$ be the closed region between the catenaries $c(g,s_g)$ and $c(g,\min(L,v_g))$. Since any two catenaries in $\{c(g,\ell):\ell\in [s_g,\min(L,v_g)]\}$ already have the points $G$ and $A$ in common, their portions between $G$ and $A$ are disjoint by Lemma \ref{lem:unique_catenary_3}. Then $C_g$ is continuously swept by the catenaries $c(g,\ell)$ as $\ell$ moves along the interval $[s_g,\min(L,v_g)]$.
\end{proof}
\subsection{Obstruction by a single point}
Consider a point $P\in\Delta$. We are interested in the set $P^-$ in the visibility region $ V_L$ obstructed by $P$, namely $P^-=\{(g,\ell):P\in c(g,\ell)\}$. We prove that $P^-$ is a continuous 
function, which in turn means that the point $P\in \Delta$ is represented by a pseudosegment $P^-\in V_L$ that connects the upper and lower boundaries of the visibility region $V_L$.

\begin{figure}[htb]
    \centering
    \includegraphics[width=0.9\linewidth]{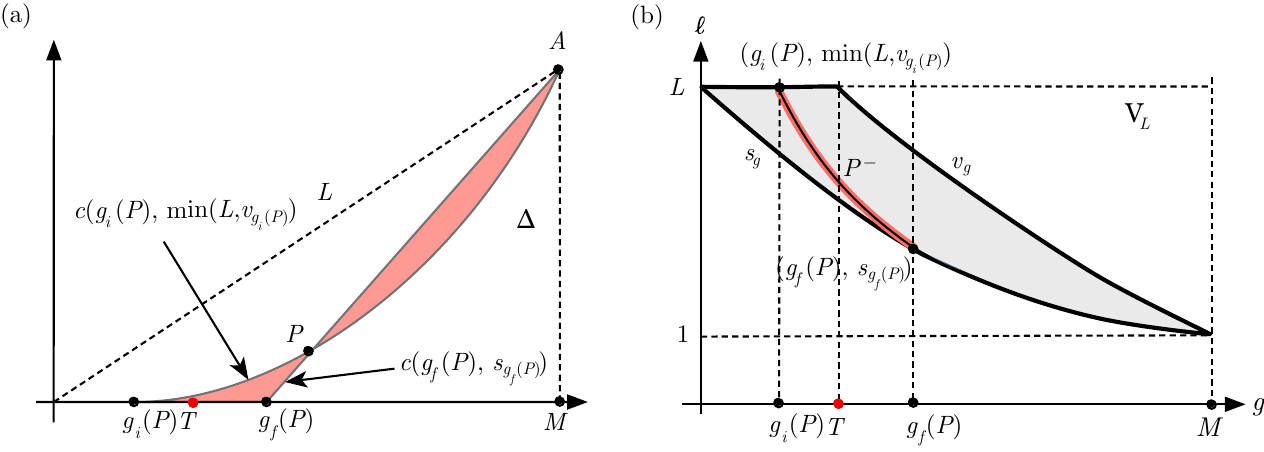}
    \caption{The visibility region $V_L$ when a point $P$ is the only obstacle in $\Delta$. The curve $P^-$ represents the set of catenaries obstructed by the point $P$. 
    }
    \label{fig:visibility2}
\end{figure}

As a consequence of Lemma \ref{lem:unique_catenary_3} we have the following Corollary.

\begin{corollary}\label{cor:curves_functions}
    Consider a point $P\in \Delta$. For every $0\leq g \leq \sqrt{L^2-1}$ there is at most one catenary $c(g,\ell)$ such that $P\in c(g,\ell)$.
\end{corollary}
 
By Corollary \ref{cor:curves_functions}, $P^-$ is a function and so we can use the notation $P^-(g)=\ell$, for the catenaries $c(g,\ell)$ passing through $P$. The function $P^-$ is continuous since such a set of catenaries continuously varies from the largest catenary to the smallest catenary $c(g,\ell)$ through $P$, see Figure \ref{fig:visibility2}. More precisely, 
\begin{equation*}
    P^-:[g_i(P),g_f(P)]\rightarrow [s_{g_f(P)},\min(v_{g_i(P)},L)]
\end{equation*}
is a continuous function where
\begin{align*}
    g_i(P)&=\min\{g:P\in c(g,\ell) \text{ for some } \ell\} \text{\hspace{.1in} and \hspace{.1in}} P^-(g_i(P)) =\min(v_{g_i(P)},L),\\
    g_f(P)&=\max\{g:P\in c(g,\ell) \text{ for some } \ell\} \text{\hspace{.1in} and \hspace{.1in}} P^-(g_f(P)) =s_{g_f(P)}.\\
\end{align*}

\subsection{Obstruction by an orthogonal rectangle}

In this section, we determine the region $R^-$ obstructed by a single \emph{orthogonal rectangle} $R$ (i.e., a rectangle with vertical and horizontal sides). This also includes the limit case where $R$ is a horizontal or vertical segment. We need the following lemmas.

\begin{lemma}\label{lem:pseudosegment}
    If $P_1$ and $P_2$ are two different points in $\Delta$, then the curves $P_1^-$ and $P_2^-$ intersect at most once.
\end{lemma}

\begin{proof}
    Suppose that these curves intersect at $(g_1,\ell_1)$ and $(g_2,\ell_2)$. Then the catenaries $c(g_1,\ell_1)$ and $c(g_2,\ell_2)$ pass through $P_1$, $P_2$, and $A$. Hence, by Lemma \ref{lem:unique_catenary_3}, the two catenaries coincide and thus $g_1=g_2$. Note that $\ell_1=\ell_2$ as they are the length of the common catenary between $(g_1,0)=(g_2,0)$ and $A$.
\end{proof}

\begin{lemma}\label{lem:disjoint_curves}
    If $P_1$ and $P_2$ are two different points in $\Delta$ such that the segment $P_1P_2$ has non-positive slope, then $P_1^-$ and $P_2^-$ do not intersect.
\end{lemma}

\begin{proof}
    If $P_1^-$ and $P_2^-$ intersected at a point $(g,\ell)\in V_L$, then the catenary $c(g,\ell)$ would pass through $P_1$ and $P_2$. This is impossible when the segment $P_1P_2$ is vertical because $c(g,\ell)$ is a function. Similarly, the portion of the catenary $c(g,\ell)$ representing the tether is to the right of its vertex. Then this portion is increasing and so it cannot pass through both $P_1$ and $P_2$.
\end{proof}

\begin{figure}[htb]
    \centering
    \includegraphics[width=01\linewidth]{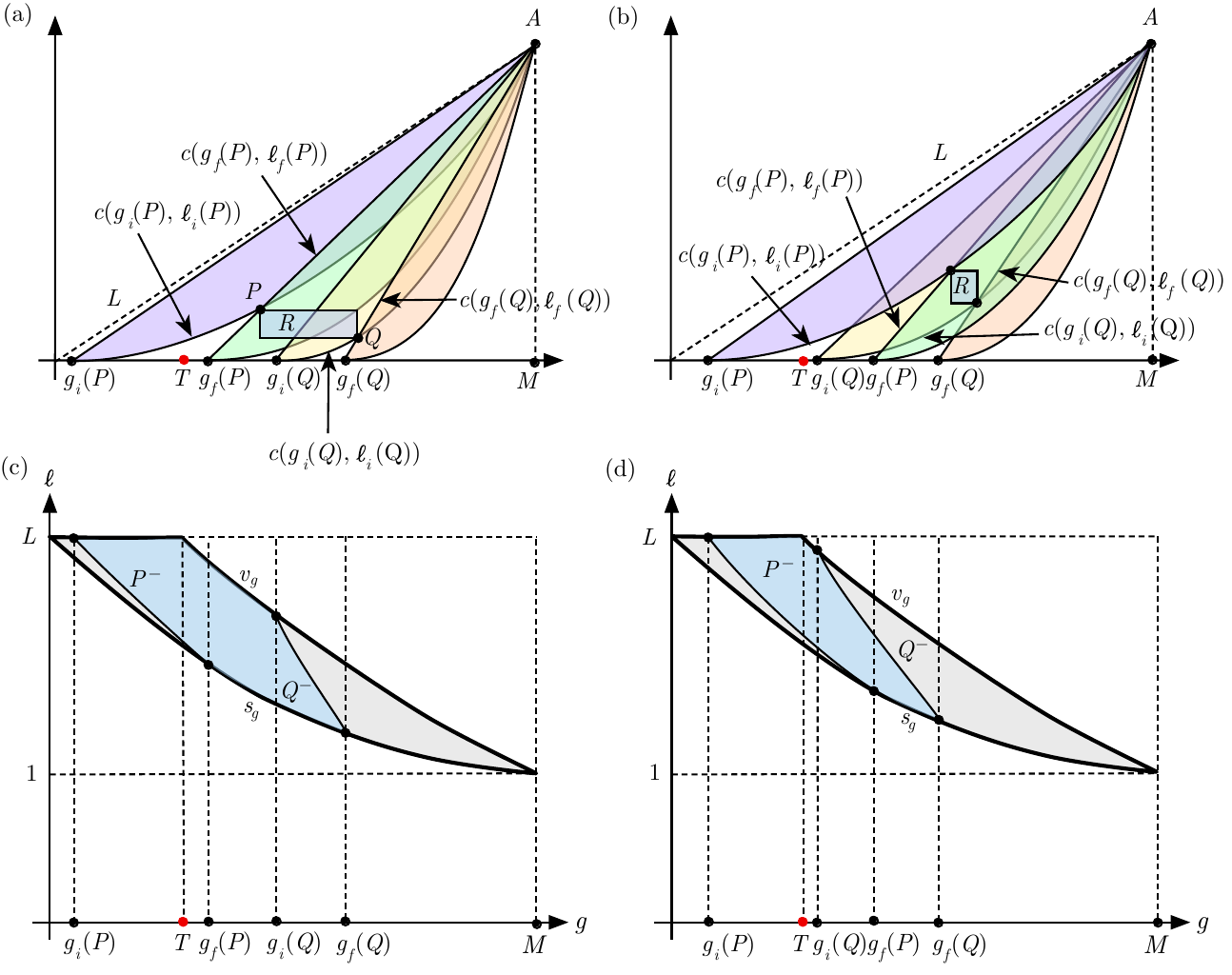}
    \caption{The visibility region when an orthogonal rectangle $R$ is the only obstacle. Two essentially different situations may occur:  (a) and (c) the shadow of $R$ is not empty when $g_i(P)\leq g_f(P)\leq g_i(Q)\leq g_f(Q)$. (b) and (d) The shadow of $R$ is empty if $g_i(P)\leq g_i(Q)\leq g_f(P)\leq g_f(Q)$. }
    \label{fig:one-rectangle}
\end{figure}

\begin{theorem}\label{th:one_rectangle}
    Let $R\subset \Delta$ be a (solid) orthogonal rectangle with upper left and lower right corners $P$ and $Q$, respectively. The set $R^-$ obstructed by $R$ is the set of all points in $V_L$ between $P^-$ and $Q^-$.
\end{theorem}
\begin{proof}
    Let $R^-\subseteq V_L$ be the region obstructed by $R$. Then $P^-\subseteq R^-$ and $Q^-\subseteq R^-$. These curves are disjoin by Lemma \ref{lem:disjoint_curves} and the end point $(g_f(P), s_{g_f(P)})$ of $P^-$  is to the left of the end point $(g_f(Q), s_{g_f(Q)})$ of $Q^-$ because the segment $PQ$ has nonpositive slope. Thus $P^-$ is completely to the left of $Q^-$ within $V_L$. Consider any other point $U\in R$. Note that $PU$ and $UQ$ are both segments with nonpositive slope and so the previous argument holds for them. That is, $P^-$ is completely to the left of $U^-$ and $U^-$ is completely to the left of $Q^-$. In other words, $U^-$ is between $P^-$ and $Q^-$. Moreover, by continuity, if we move a point $U$ from $P$ to $Q$ following a decreasing curve, then the corresponding curves $U^-$ sweep the region enclosed between $P^-$ and $Q^-$, proving the statement. 
\end{proof}

\begin{proposition}
     Let $R\subset \Delta$ be a (solid) orthogonal rectangle with upper left corner $P$ and lower right corner $Q$. If $ g_f(P)\leq g_i(Q)$, then $R$ has shadow $[g_f(P),g_i(Q)]$. If $g_i(Q)< g_f(P)$, then the shadow of $R$ is empty. In both cases, if $R$ is the only obstacle in $\Delta$, there is full visibility for $g<g_i(P)$ as well as for $g>g_f(Q)$.
\end{proposition}
\begin{proof}
  As we shown in the proof of Theorem \ref{th:one_rectangle} $P^-$ is completely to the left of $Q^-$, therefore, the only possible orders for these four parameters are $g_i(P)\leq g_f(P)\leq g_i(Q)\leq g_f(Q)$ and $g_i(P)\leq g_i(Q)\leq g_f(P)\leq g_f(Q)$, see Figures \ref{fig:one-rectangle} (a-b). The corresponding region $R^-$ obstructed by $R$ is shown in Figures \ref{fig:one-rectangle} (c-d). In both cases, there is full visibility from any $g\in[0,g_i(P))\cup(g_f(Q),\sqrt{L^2-1}]$. Note that, when $g_f(P)\leq g_i(Q)$ there is null visibility  whenever  $g_f(P)\leq g \leq g_i(Q)$, and partial visibility for any other location $g\in(g_i(P),g_f(P))\cup(g_i(Q),g_f(Q))$, Figure \ref{fig:one-rectangle} (c). If $g_i(Q)\leq g_f(P)$, there is partial visibility from every position $g\in [g_i(P),g_f(Q)]$, Figure \ref{fig:one-rectangle} (d).
\end{proof}

\subsection{The visibility region when the obstacles are orthogonal rectangles}
In this section, we analyze the case where the obstacles are all orthogonal rectangles. This case is useful in practice, as any finite set of obstacles (not necessarily rectangular) can be associated with a family of orthogonal rectangles by enclosing each obstacle in the smallest possible orthogonal rectangle. The visibility region of the original family of obstacles contains the visibility region of the associated family of rectangular obstacles. 

\begin{figure}[htb]
    \centering
    \includegraphics[width=.5\linewidth]{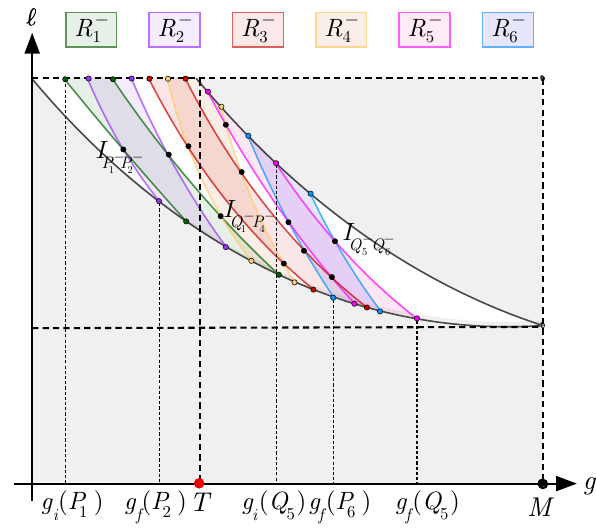}
    \caption{The arrangement $\cal{A(\widetilde{\cal{R}})}$  for a set of obstacles ${\cal R}=\{R_1,R_2,R_3,R_4,R_5,R_6\}$.}
\label{fig:celdas}
\end{figure}

\begin{theorem}\label{th:fam_pseudoseg}
    Let $\mathcal{R}=\{R_1, R_2, \dots, R_n\}$ be a collection of $n$ disjoint orthogonal rectangles in $\Delta$. For $1\leq i\leq n$, let $P_i$ and $Q_i$ be the upper left and lower right corners of $R_i$, respectively. Then the visibility region for $\mathcal{R}$ is the set $\mathcal{S}:=V_L\setminus \mathcal{R}^-$, where $\mathcal{R}^-=\bigcup_{i=1}^n R_i^-$ and $R_i^-$ is the subset of $V_L$ between $P_i^-$ and $Q_i^-$. Moreover, the set 
    \[
    \widetilde{\mathcal{R}}:=\{P_i^-:1\leq i\leq n\}\cup \{Q_i^-:1\leq i\leq n\}\cup\{\beta_1,\beta_2,\beta_3\}
    \]
    is a family of $2n+3$ pseudosegments contained in $V_L$, where $\beta_i$, $i=1,2,3,$ are the curves defined in Proposition \ref{prop:V_L}.
\end{theorem}    
\begin{proof}
    The first part of the statement follows directly from Theorem \ref{th:one_rectangle} applied to each rectangle $R_i$. For the second part, each element of  $\widetilde{\mathcal{R}}$ is a simple continuous curve in $V_L$; and any two curves $S$ and $S'$ in $\widetilde{\mathcal{R}}$ intersect at most once, by Lemma \ref{lem:pseudosegment} and Proposition \ref{prop:V_L}. In other words, $\widetilde{\mathcal{R}}$ is an arrangement of pseudosegments. 
\end{proof}
Given a set $\mathcal{R}=\{R_1, R_2, \dots, R_n\}$ of disjoint orthogonal rectangles, let $\cal{A}(\widetilde{\cal{R}})$ be the arrangement determined by the family of pseudosegments $\widetilde{\mathcal{R}}$ defined in Theorem \ref{th:fam_pseudoseg} (see Figure \ref{fig:celdas}). For $S\neq S'$ in $\widetilde{\mathcal{R}}$, let $I_{SS'}$ be the intersection of the pseudosegments $S$ and $S'$, if it exists. These points are precisely the vertices of $\cal{A}(\widetilde{\cal{R}})$. 
Note that for every $P$ in $\Delta$, $(g_f(P),s_{g_f(P)})=P^-\cap \beta_1$ and $(g_i(P),\min(L,v_{g_i(P)}))=P^-\cap (\beta_2\cup \beta_3)$.

\subsection{A visibility range with a quadratic number of connected components}
In this section, we describe a set of $n$ obstacles such that its visibility region has a quadratic number of connected components with disjoint vertical projections. This gives a natural $\Omega(n^2)$ time lower bound for the 2D-CV problem.

\begin{theorem}\label{th:2d_example}
    There is a family $\mathcal{R}_n$ of $n$ orthogonal rectangular obstacles, whose visibility region $V_L\setminus\mathcal{R}_n^-$ has $\Theta(n^2)$ vertices. Moreover, the visibility range of $\mathcal{R}_n$, i.e., the projection of $V_L\setminus\mathcal{R}_n^-$ onto the $g$-axis, consists of $\Theta(n^2)$ disjoint closed intervals.  
\end{theorem}

\begin{figure}
    \centering
    \includegraphics[width=.9
\linewidth]{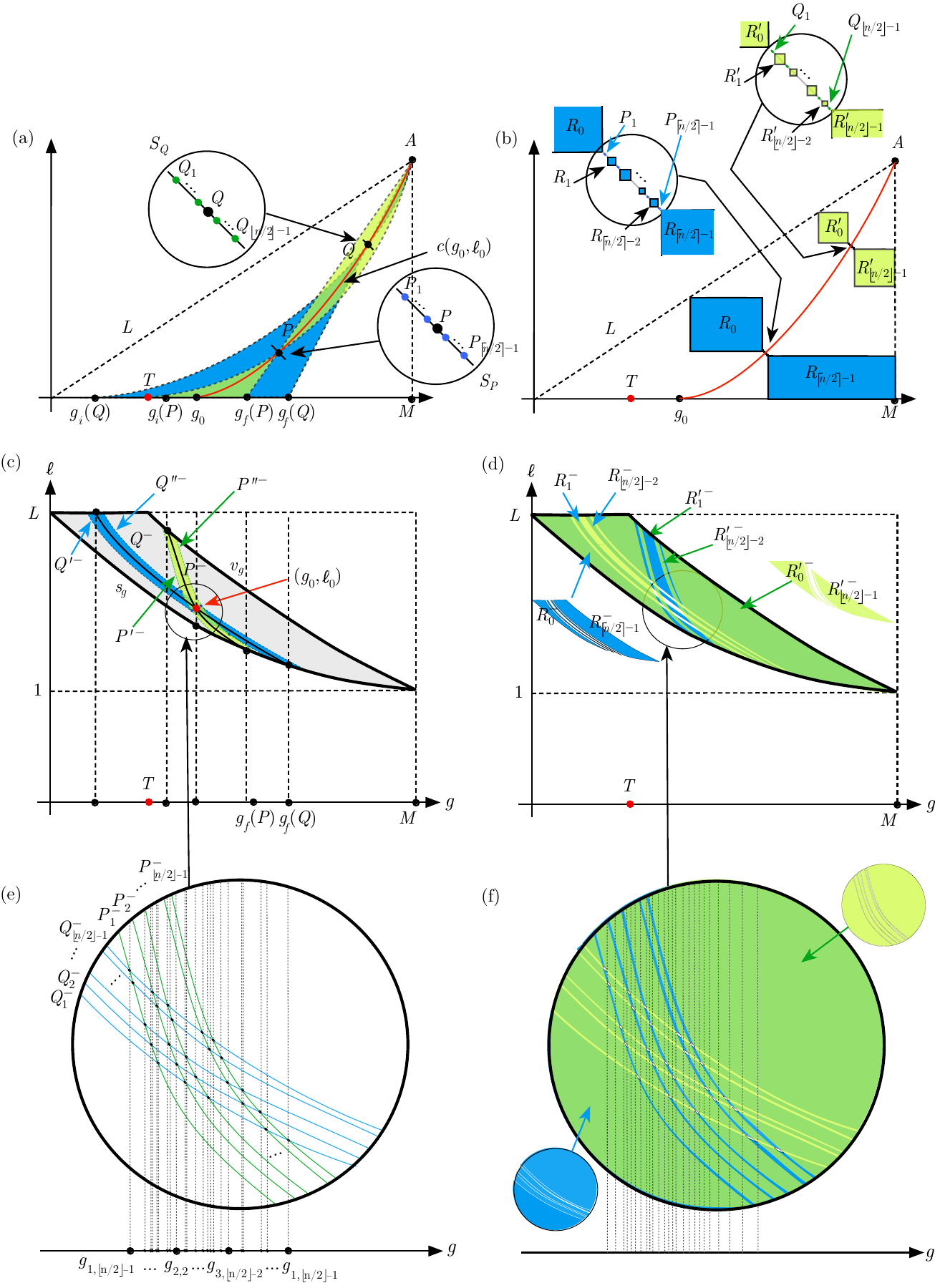}
    \caption{A family of obstacles whose visibility region has a quadratic number of connected components. }
    \label{fig:example_quad}
\end{figure}

\begin{proof}
   We construct the family of $n$ obstacles $\mathcal{R}_n=\{R_0, R_1, \dots,R_{\lceil n/2 \rceil-1},R'_0,R'_1,\dots,R'_{\lfloor n/2 \rfloor-1}\}$, all orthogonal rectangles. Consider any catenary $c(g_0,\ell_0)$ such that $s_{g_0}<\ell_0<\min(v_{g_0},L)$, and let $P$ and $Q$ be two points in $c(g_0,\ell_0)$. This guarantees that the point $(g_0,\ell_0)$ is in the interior of $V_L$ and is the intersection of $P^-$ and $Q^-$.  Consider two small segments $S_P=P'P''$ and $S_Q=Q'Q''$ with slope $-1$ passing through $P$ and $Q$, respectively (see Figure \ref{fig:example_quad}a), such that $S_P$ is contained in the region defined by all catenaries $c(g,\ell)$ passing through $Q$; and $S_Q$ is contained in the region defined by all catenaries $c(g,\ell)$ passing through $P$. This means that $S_P^-\cap S_Q^-$ is completely contained in the interior of $V_L$ (see Figure \ref{fig:example_quad}c).

   Consider the points, $P_1,P_2,\dots,P_{\lceil n/2 \rceil-1}$ in $S_P$ and $Q_1,Q_2,\dots,Q_{\lfloor n/2 \rfloor-1}$  in $S_Q$, such that the catenaries through $P_i$ and $Q_j$ have different $x$-intercepts $g_{i,j}$. By Theorem \ref{th:one_decrcurve} applied to the segments  $S_P$ and $S_Q$, this is equivalent to selecting the curves $P_1^-,P_2^-,\dots,P_{\lceil n/2 \rceil-1}^-$ in $S_P^-$ and the curves $Q_1^-,Q_2^-,\dots,Q_{\lfloor n/2 \rfloor-1}^-$ in $S_Q^-$ so that all their $(\lceil n/2 \rceil-1)(\lfloor n/2 \rfloor-1)$ points of intersection have different $g$-coordinates (see Figure \ref{fig:example_quad}e).
   
   Let $\epsilon>0$ small enough such that the intervals $(g_{i,j}-\epsilon,g_{i,j}+\epsilon)$ do not intersect.  With center at $P_i$ and $Q_j$, consider segments $S_i=P_i'P_i''\subset S_P$ and $S'_j=Q_j'Q_j''\subset S_Q$ of length $2\delta$ for some $\delta>0$ small enough such that for any catenary $c(g,\ell)$ that intersects $S_i$ and $S'_j$ it holds that $g\in(g_{i,j}-\epsilon,g_{i,j}+\epsilon)$. This is indeed possible as in $V_L$, these catenaries correspond to the intersection of the (open) regions $S_i^-$ (bounded by $P'^-_i$ and $P''^-_i$) and ${S'_j}^-$ ($Q'^-_j$ and $Q''^-_j$). By Theorem  \ref{th:one_decrcurve}, these regions get thinner as $\delta$ decreases and their intersection converges to the point $P_i^-\cap Q_j^-$. So for $\delta$ small enough, all projections of $S_i^-\cap {S'_j}^-$ will be contained in $(g_{i,j}-\epsilon,g_{i,j}+\epsilon)$ as desired.
   
   We refer to the segments $S_i$ with $1\leq i \leq \lceil n/2 \rceil-1$ and $S'_j$ with $1\leq j \leq \lfloor n/2 \rfloor-1$ as the $P$- and \emph{$Q$-windows of visibility}, respectively.  We now describe the rectangles, which are essentially generated by the remaining segments contained in  $S_P$ and $S_Q$ when the windows of visibility are removed. For each $1\leq i <\lceil n/2 \rceil-2$, let $R_i$ be the unique square having $P''_i$ and $P'_{i+1}$ as vertices (see Figure \ref{fig:example_quad}b); and for each $1\leq j <\lfloor n/2 \rfloor-2$, let $R'_j$ be the unique square having $Q''_j$ and $Q'_{j+1}$ as vertices. Finally, let $R_0$ and $R'_0$ be orthogonal rectangles whose lower right corner is $P'_1$ and $Q'_1$, respectively, and upper left corners are in $OA$; and let $R_{\lceil n/2 \rceil}$ and $R'_{\lfloor n/2 \rfloor}$ be rectangles whose upper left corner is $P''_{\lceil n/2 \rceil-1}$ and $Q''_{\lfloor n/2 \rfloor-1}$, respectively, and lower right corners are in $MA$.
    
   We now argue that the visibility region $V_L\setminus\mathcal{R}_n^-$ of $\mathcal{R}_n$,  has $\Theta(n^2)$ vertices. In fact, we first prove  that the projection of such visibility region onto the $g$-axis consists of $\Theta(n^2)$ disjoint closed intervals. 
   
   In $V_L$ (see Figure \ref{fig:example_quad}c), the  regions $R_i^-$ and ${R'}_j^-$ are bounded by $P''^-_i$ and $P'^-_{i+1}$ for $i\neq 0$ and $i\neq \lceil n/2 \rceil-1$ and by $Q''^-_j$ and $Q'^-_{j+1}$ for $j\neq 0$ and $j\neq\lfloor n/2 \rfloor-1$, respectively. The regions $R_0^-$ and ${R'}_0^-$ are bounded by $P'^-_1$ and $Q'^-_1$ to their right, respectively, and by the boundary of $V_L$ above and below. Similarly, $R_{\lceil n/2 \rceil-1}^-$ and ${R'}_{\lfloor n/2 \rfloor-1}^-$ are bounded by $P'^-_{\lceil n/2 \rceil-1}$ and $Q'^-_{\lfloor n/2 \rfloor-1}$ to their left, respectively, and by the boundary of $V_L$ above and below. So $V_L$ is covered by $\bigcup_{i=0}^{\lceil n/2 \rceil-1}R_i^-$ except for the thin regions  $S_i^-$ surrounding $P^-_i$; and simultaneously $V_L$ is covered by $\bigcup_{j=0}^{\lfloor n/2 \rfloor-1}{R'}_j^-$ except for the thin regions  ${S'}_j^-$ surrounding $Q_j^-$. So $$V_L\setminus\mathcal{R}_n^-=\bigcup \{S_i^-\cap{S'}_j^-:1\leq i\leq \lceil n/2 \rceil-1 , 1\leq j\leq \lfloor n/2 \rfloor-1\}.$$
   
   Since the projection of $S_i^-\cap{S'}_j^-$ onto the $g$-axis is a closed interval contained in $(g_{i,j}-\epsilon,g_{i,j}+\epsilon)$ and all these last open intervals are disjoint by construction, then the projection of $V_L\setminus\mathcal{R}_n^-$ onto the $g$-axis consists of the disjoint union of exactly $(\lceil n/2 \rceil-1)(\lfloor n/2 \rfloor-1)=\Theta (n^2)$ closed intervals (see Figure \ref{fig:example_quad}f). Moreover, this implies that the regions $S_i^-\cap{S'}_j^-$ are disjoint, each bounded by four of the pseudosegments in $\widetilde{\mathcal{R}}_n$ and so its vertices are 
   $P'^-_i\cap Q'^-_j, P'^-_i\cap Q''^-_j,
   P''^-_i\cap Q'^-_j$ and $P''^-_i\cap Q''^-_j$; that is, $\mathcal{R}_n^-$ has exactly $4(\lceil n/2 \rceil-1)(\lfloor n/2 \rfloor-1)=\Theta (n^2)$ vertices.
\end{proof}

\subsection{Computing the Catenary Visibility Range}\label{subsect:computing_visibility_range}

As for any computational problem, the model of computation on which to solve the problem must be stated. 
For the computational problems of this paper,  the computational model is the Real RAM model, augmented with constant-time primitives for calculating the intersection point of a pair of curves of our arrangements, as well as for calculating the intersection of one of these curves with one given vertical or horizontal line. Each of these two operations requires solution of a numerical problem that involves implicit equations with hyperbolic trigonometric functions. We assume that the required numerical precision is fixed throughout the computation, so that these operations can be implemented by numerical solver routines with constant running time in the standard Real RAM model. All running-time bounds in this paper are understood with respect to this augmented model.

Let $\mathcal{R}=\{R_1,R_2,\dots,R_n\}$ be a set of $n$ orthogonal rectangular obstacles corresponding to an instance of the 2D-CV problem. Consider the corresponding family of pseudosegments $\widetilde{\mathcal{R}}$ and the arrangement $\mathcal{A(\widetilde{\mathcal{R}})}$ in the visibility plane. 
An example of such an arrangement of pseudosegment is illustrated in Figure \ref{fig:celdas}. Note that there are (at most) five curves that bound each region $R^-_i$ obstructed by a rectangle $R_i$, namely $P^-_i$, $Q^-_i$, $\ell=s_g$, $\ell=v_g$, and $\ell=L$. The \emph{depth} of a cell of $\mathcal{A(\widetilde{\mathcal{R}})}$ is the number of obstructed regions $R^-_i$ that contain it. 
The following proposition follows trivially from the definition of the region of an obstacle.

\begin{proposition}\label{prop-zero}
The union of zero-depth cells in $\mathcal{A(\widetilde{\mathcal{R}})}$ corresponds to the set of collision-free catenaries $c(g,\ell)$, i.e., the visibility region of $\cal R$. Thus, the solution to the 2D-CV  problem is given by the vertical projection of the union of  the zero-depth cells in $V_L$ to the $g$-axis, which defines a set of intervals within $[O,M]$. The rest of $[O,M]$ is the shadow of $\mathcal{R}$.
\end{proposition}

Given the target point, the length of the tether (note that we normalize so that the target point $A$ is at distance 1 from the ground and $L$ from the origin) and the set of rectangular obstacles, we want to compute the set $I = \{I_1, \ldots, I_m\}$ of subintervals of $[O, M]$, whose union is the set of positions $g$ for which there exists a collision-free catenary $c(g, \ell)$. By Proposition \ref{prop-zero}, this union of intervals corresponds to the vertical projection of the cells with depth zero in $V_L$.

\vspace{2mm}

Our procedure can be derived as a simple modification of the well known Bentley-Ottmann algorithm for general curves~\cite{BentleyOttmann}. 
This algorithm has as input a set of curves, 
and visits all the intersection points induced by them,  using a left-to-right sweep line, recording the relative bottom-up ordering of the curves as they are intersected by the sweep line.
The only modification we need is to calculate the depth of each  intersected cell. This can be done using the following observations: (1) The cells intersected by the sweep line, their depths, and their bottom-up order do not change as long as the sweep line does not encounter a vertex of the arrangement.
(2) When the sweep line crosses an intersection point, only a constant number of intersected cells is either added or removed, or changes depth, and in all cases, such cell is adjacent to the processed intersection point, i.e. adjacent to the intersecting curves.  The Bentley-Ottman data structure allows us to identify, in constant time, which of the curves intersect at the processing point. In turn, this allows us to locate which of the cells to insert/remove/updateDepth in constant time. 

The running time of the Bentley-Ottman algorithm is $O((n+k)\log n)$, where $n$ is the number of obstacles, and $k$ is the number of intersections among the curves (i.e., the vertices of $\mathcal{A(\widetilde{\mathcal{R}})}$), and it can be implemented with $O(n)$ working memory~\cite{Brown1981}. Since in our context two curves in the visibility region intersect at most once (Lemma  \ref{lem:pseudosegment}), the number $k$ of intersections among them could be at most $O(n^2)$. Thus, we arrive to the following result:
\begin{theorem}\label{th:CVPAlgorithm}
The 2D-CV problem can be solved in $O((n+k) \log n)$, where $n$ is the number of rectangular obstacles and $k$ is the number of vertices in the arrangement $\mathcal{A(\widetilde{\mathcal{R}})}$.
\end{theorem}

\subsection{Reporting feasible catenary lengths from a launch point }

\begin{figure}
    \centering
    \includegraphics[width=.5\linewidth]{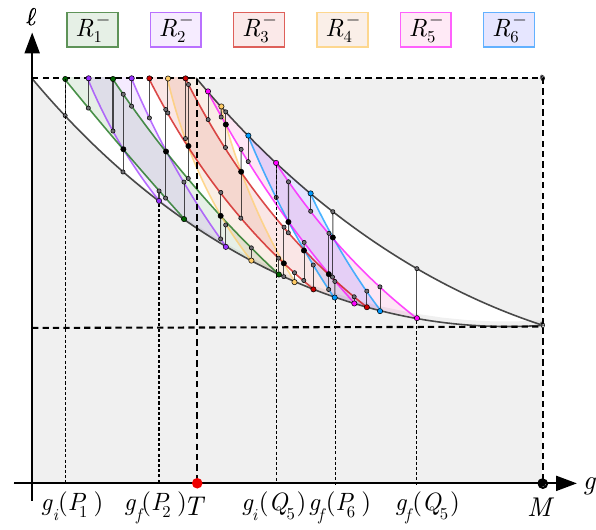}
    \caption{The vertical decomposition (trapezoidal map) of the arrangement $\cal{A(\widetilde{\cal{R}})}$ in Figure \ref{fig:celdas}, for a set of obstacles $\widetilde{\cal{R}}=\{R_1,R_2,R_3,R_4,R_5,R_6\}$.}
\label{fig:celdasTrap}
\end{figure}

For safe operation in marsupial systems, the tether must maintain a minimum clearance from obstacles, not merely avoid collision. Consequently, designing robust trajectories requires knowing a range of admissible tether lengths. This allows paths to accommodate small variations in tether deployment while preserving the required safety margin. To address this issue, this section presents an efficient algorithm for answering the following query problem:

\emph{Given a set of orthogonal rectangular obstacles $\mathcal{R}$ and a launch location $(g,0)$, report the set of feasible lengths $F_g$, that is, all tether lengths $\ell$ for which the catenary $c(g,\ell)$ is collision-free.
}

In our context, this question translates into the following problem: Given a value $0\leq g\leq \sqrt{L^2-1}$, obtain the intersection of the union of the cells of depth zero in the arrangement $\mathcal{A(\widetilde{\mathcal{R}})}$, with the vertical line passing through $(g,0)$ in the visibility plane.

Observe that since each curve of the arrangement $\mathcal{A(\widetilde{\mathcal{R}})}$ can be intersected by a given vertical line in at most one point, then the intersection of such a vertical line with a cell of the arrangement $\mathcal{A(\widetilde{\mathcal{R}})}$ is a (possibly empty) vertical segment. Thus, the natural output of the query is a list of disjoint intervals of ordinates of the visibility plane; that is, one such interval per intersected cell of depth zero.

For a given set of orthogonal rectangular obstacles  $\mathcal{R}$ and a launch location $(g,0)$, let $V$ be the vertical line passing through  $(g,0)$ in the visibility plane, and let $m\leq 2n+3$ be the number of curves of the arrangement $\mathcal{A(\widetilde{\mathcal{R}})}$ intersected by $V$. Let $C_1,\ldots,C_m$ be those intersected curves, sorted by increasing ordinate value of its intersection point with $V$. Denote as $p_1\ldots,p_m$ the vertically ascending sequence of those intersection points (suppose w.l.o.g. that all the intersection points are distinct). Observe that the vertically ascending intersections of $V$ with the cells of $\mathcal{A(\widetilde{\mathcal{R}})}$, is precisely the sequence of vertical segments from one intersection point to the next in the sequence $p_1,\ldots,p_m$. 

Under the assumption that no two curves are intersected by $V$ in the same point, the depth of the cell corresponding to the first segment has depth zero. 
Note that for any two consecutive segments, $\overline{p_ip_{i+1}}$ and $\overline{p_{i+1}p_{i+2}}$ of this sequence, the corresponding cells have depths that differ in exactly one unit: the depth of the upper cell is larger (resp. smaller) by one unit if the point $p_{i+1}$ corresponds to the intersection of $V$ with a curve induced by the lower-left (resp. upper-right) point of an obstacle $R$ in  $\mathcal{R}$ (intuitively, when going from the lower cell to the upper cell, we \emph{enter} (resp. \emph{exit}) the region $R^{-}$ obstructed by $R$, and only that region.

The previous observation reduces the query problem to the following algorithm: 1) identify which curves of $\mathcal{A(\widetilde{\mathcal{R}})}$ are intersected by the line $V$; 
2) calculate and sort vertically the ordinate values of the intersected curves with $V$; 3) visit each consecutive pair of intersection points, updating the depth of the corresponding cell, according to the type of the curve associated to the intermediate point; 4) report those pair of points that correspond to cells of depth zero.

The step 1 of the previous algorithm can be trivially done in $O(n)$ time, observing that the vertical line through the point $(g,0)$ intersects the two curves associated to an obstacle $R$ if and only if $g$ is contained in the interval $(g_i(R),f_i(R))$. The step 2 of the algorithm involves the calculation of the $m$ intersection points, as well as time $O(m\log m)$ to vertically sort them. Steps 3 and 4 take time $O(m)$. Thus, the total running time for answering a query is $O(n + m\log m)$.

It is natural to consider that for a given fixed set of rectangular orthogonal obstacles, it might be desirable to answer the query problem for several distinct vertical query lines. In this case, we can opt for preprocessing the given set of input rectangles into a data structure, in order to achive a lower query time when the number $m$ of intersected curves of the arrangement is asymptotically smaller than the total number $n$ of obstacles. Note that the only step of the former algorithm that has running time proportional to $n$ is step 1. Thus, we optimize this step. This is a well known task that can be carried out with the use of \emph{interval trees} \cite{de2008computational}. This is a data structure that processes $n$ intervals in time $O(n \log n)$ using $O(n)$ space, and given a value $x$, reports all the intervals that contain the value $x$ (let $m$ be this number of intervals), in time $O(\log n + m )$. With the addition of this data structure and  paying this $O(n \log n)$ preprocessing time and $O(n)$ space, step 1 of our algorithm now takes time $O(\log n + m)$, and thus the total running time for answering a query can be lowered to $O(\log n + m \log m)$, which can be expressed more concisely as $O(m \log n)$, where $m$ is the number of curves of the arrangement $\mathcal{A(\widetilde{\mathcal{R}})}$ intersected by the vertical query line $V$.

\begin{theorem}\label{th:query}
The query problem can be solved in $O(m \log n)$, where $n$ is the number of rectangular
obstacles and $m$ denotes the size of the output.
\end{theorem}

\section{The 3-dimensional scenario}
\begin{figure}
    \centering
    \includegraphics[width=.8\linewidth]{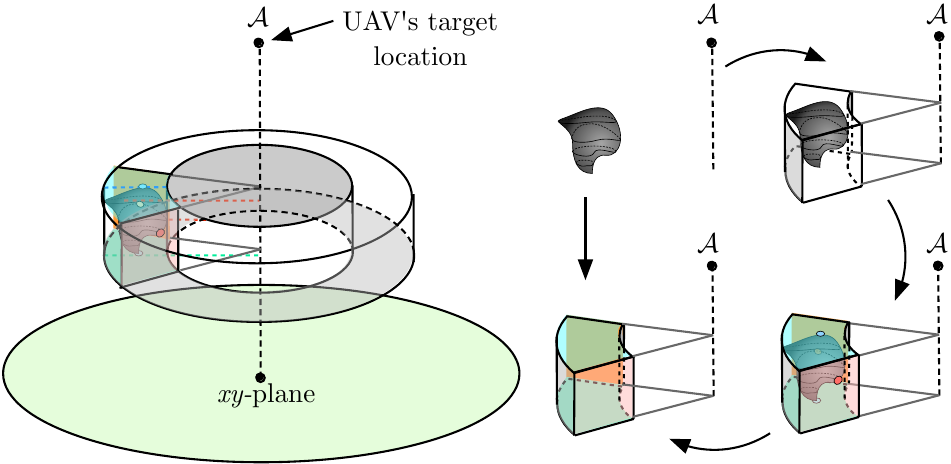}
    \caption{The smallest ACW-box containing an obstacle.}
    \label{fig:3D_boxes}
\end{figure}

Now, we consider the 3-dimensional version of the catenary visibility problem (3D-CV problem). We previously simplified the 2-dimensional problem by enclosing each obstacle within the smallest orthogonal rectangle containing it. Similarly, we simplify the 3-dimensional problem by enclosing each obstacle within the unique \emph{Annular Cylindrical Wedge},  with vertical central axes passing through the UAV's target point, which is tangent to the obstacle at each of its six two-dimensional boundaries: four planes and the lateral faces of two cylinders (see Figure \ref{fig:3D_boxes}). We refer to such enclosing objects as \emph{ACW-boxes}. 

\begin{figure}
    \centering
    \includegraphics[width=1\linewidth]{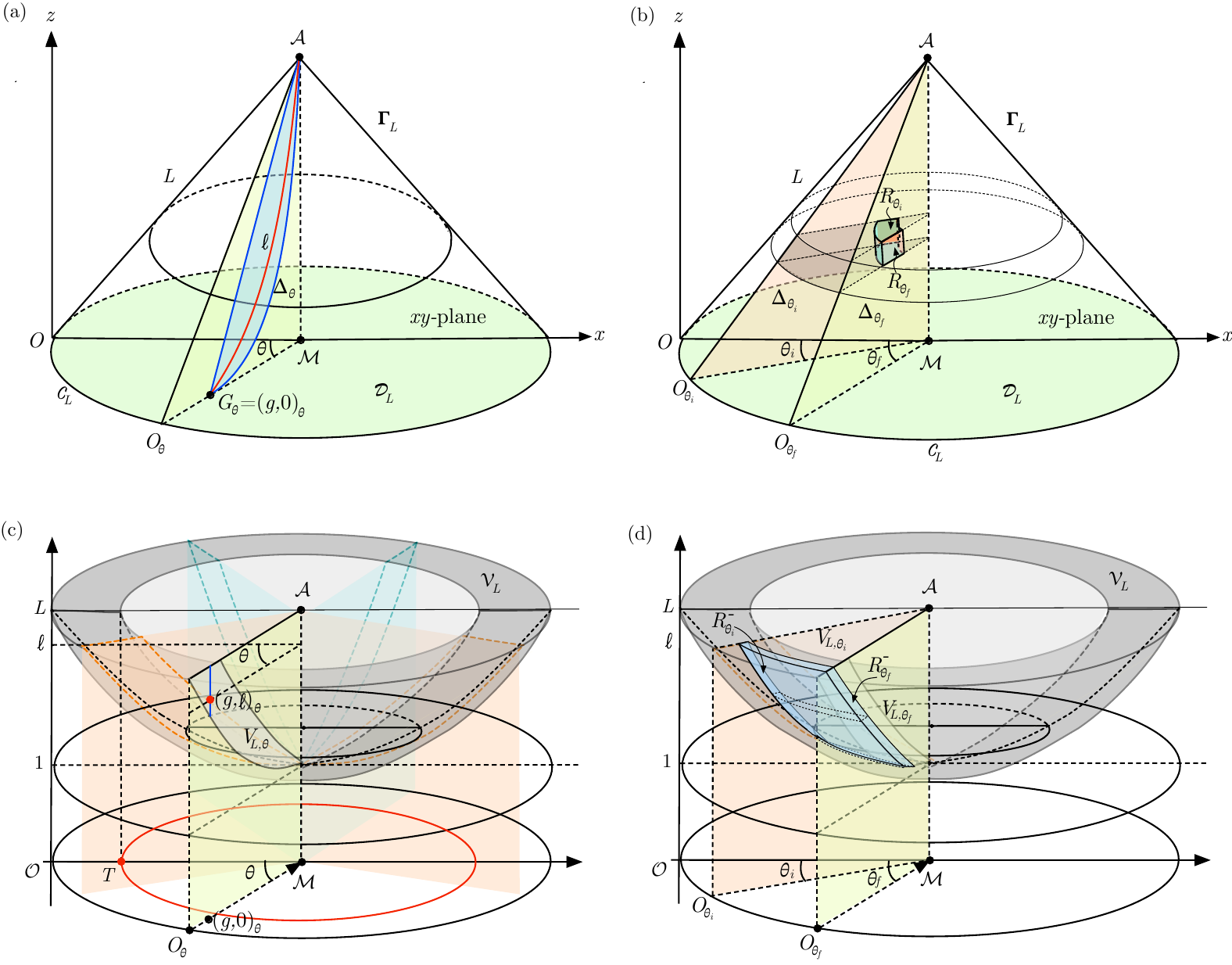}
    \caption{(a) The 3-dimensional model showing the catenary $c(g,\ell)_\theta$. (b) An ACW-box with lateral faces $R_{\theta_i}$ and $R_{\theta_f}$; (c) The visibility region showing the point $(g,\ell)_\theta$ corresponding to the catenary $c(g,\ell)_\theta$. (d) The region obstructed by the ACW-box in (b).}
    \label{fig:3d_visibility}
\end{figure}

In this 3D model, we assume that the UGV moves in the $xy$-plane, the tether has a maximum length extension $L$, and the UAV's target location is represented by the point $\mathcal{A}=(\sqrt{L^2-1},0,1)$, see Figure \ref{fig:3d_visibility}a. This means that the tether is completely extended when the UGV is at any point of circumference $\mathcal{C}_L$ on the $xy$-plane, with center $\mathcal{M}=(\sqrt{L^2-1},0,0)$ passing through the origin $\mathcal{O}=(0,0,0)$ (radius $\sqrt{L^2-1}$). Let $\mathcal{D}_L$ be the disk bounded by $\mathcal{C}_L$ and $\Gamma_L$ be the vertical cone with base $\mathcal{D}_L$ and vertex $\mathcal{A}$. For an angle $\theta\in [0,2\pi)$, let $O_\theta$ be the point in $\mathcal{C}_L$ such that $\angle \mathcal{OM}O_\theta=\theta$ and $\Delta_\theta=\triangle O_\theta \mathcal{MA}$. Fixing $\theta$ restricts the movement of the tether and UGV to $\Delta_\theta$, which corresponds to the 2-dimensional problem in $\Delta_\theta$. The 2-dimensional obstacles in $\Delta_\theta$ are the intersections of the original 3-dimensional obstacles with $\Delta_\theta$. Using this observation,  we can extend the 2-dimensional notation by keeping track of the angle $\theta$ that is being used: a point $P\in \Gamma_L$ is identified with $(x,y)_\theta$ for some $\theta\in [0,2\pi)$; $G_{\theta},=(g,0)_\theta$ represents the location of the UGV, for some $0\leq g\leq \sqrt{L^2-1}$. The visibility region $\mathcal{V}_L$, when there are no obstacles present, (see Figure \ref{fig:3d_visibility}c) is the union of the 2-dimensional visibility regions $V_{L,\theta}$ over all possible angles $\theta\in [0,2\pi)$, that is, $\mathcal{V}_L=\bigcup_{\theta\in [0,2\pi)}V_{L,\theta}$. Therefore, the point $(g,\ell)_\theta$ of the visibility region $\mathcal{V}_L$ corresponds to having the UGV at position $G_{\theta}=(g,0)_\theta$ and the tether extended to length $\ell$; the corresponding catenary described by the tether is denoted by $c( g,\ell)_\theta$. We similarly extend the notation $s_g$ and $v_g$ to $s_g(\theta)$ and $v_g(\theta)$, respectively. Figure \ref{fig:3d_visibility}b shows the ACW-box $R_{[\theta_i,\theta_f]}$ in $\Gamma_L$. This box satisfies that $R_{[\theta_i,\theta_f]}\cap\Delta_\theta$ is the copy of a rectangle $R\in\Delta$  for each $\theta_i\leq \theta \leq \theta_j$ and empty otherwise. The region $R^-_{[\theta_i,\theta_f]}$ obstructed by $R_{[\theta_i,\theta_f]}$ in $\mathcal{V}_L$ is shown in Figure \ref{fig:3d_visibility}d. 

Given a set of 3-dimensional obstacles $\mathcal{B}$, the \emph{shadow} of $\mathcal{B}$ is the set of points $G_\theta\in \mathcal{D}_L$ with no visibility from the target point $\mathcal{A}$. More precisely, the shadow of  $\mathcal{B}$ is $\{(g,0)_\theta\in \mathcal{D}_L:0\leq \theta <2\pi,F_g=\emptyset \text{ in } \Delta_\theta\}$. The output of the 3D-CV problem, called  the \emph{catenary visibility domain} of $\mathcal{B}$, is the set of points in $\mathcal{D}_L$ that are not in the shadow of  $\mathcal{B}$.

\subsection{Computing the Catenary Visibility Domain }

Observe that for a given instance with $n$ obstacles of the 3D-CV problem, the number of combinatorially  distinct instances of the 2D-CV problem (those induced by the intersection of some $\Delta_\theta$ plane with the obstacles) is at most $2n$. By Theorem \ref{th:CVPAlgorithm}, each of them can be solved independently in $O(n^2 \log n)$ time, yielding a total time of $O(n^3 \log n)$. Although quite simple, this might seem naive, because it is just natural to think that processing these $O(n)$ instances consecutively performing an angular $\theta$ sweep would yield a faster dynamic algorithm that allows reuse of the solution from one $\theta$ position to the next. Nevertheless, while it is true that the insertion (or removal) of one obstacle to a 2D instance adds (or removes) at most $n$ vertices and edges to the corresponding arrangement, it may always happen that the size of the solution (intervals of feasible drone launch parameters) changes its value by $\Theta(n^2)$.    Theorem \ref{th:3d_example} formalizes this observation, proving that no other strategy can be asymptotically faster in general (in the worst case) than that of solving the $O(n)$ 2D subinstances independently (up to possibly a logarithmic factor). 

\begin{figure}[hbt]
    \centering
    \includegraphics[width=.8\linewidth]{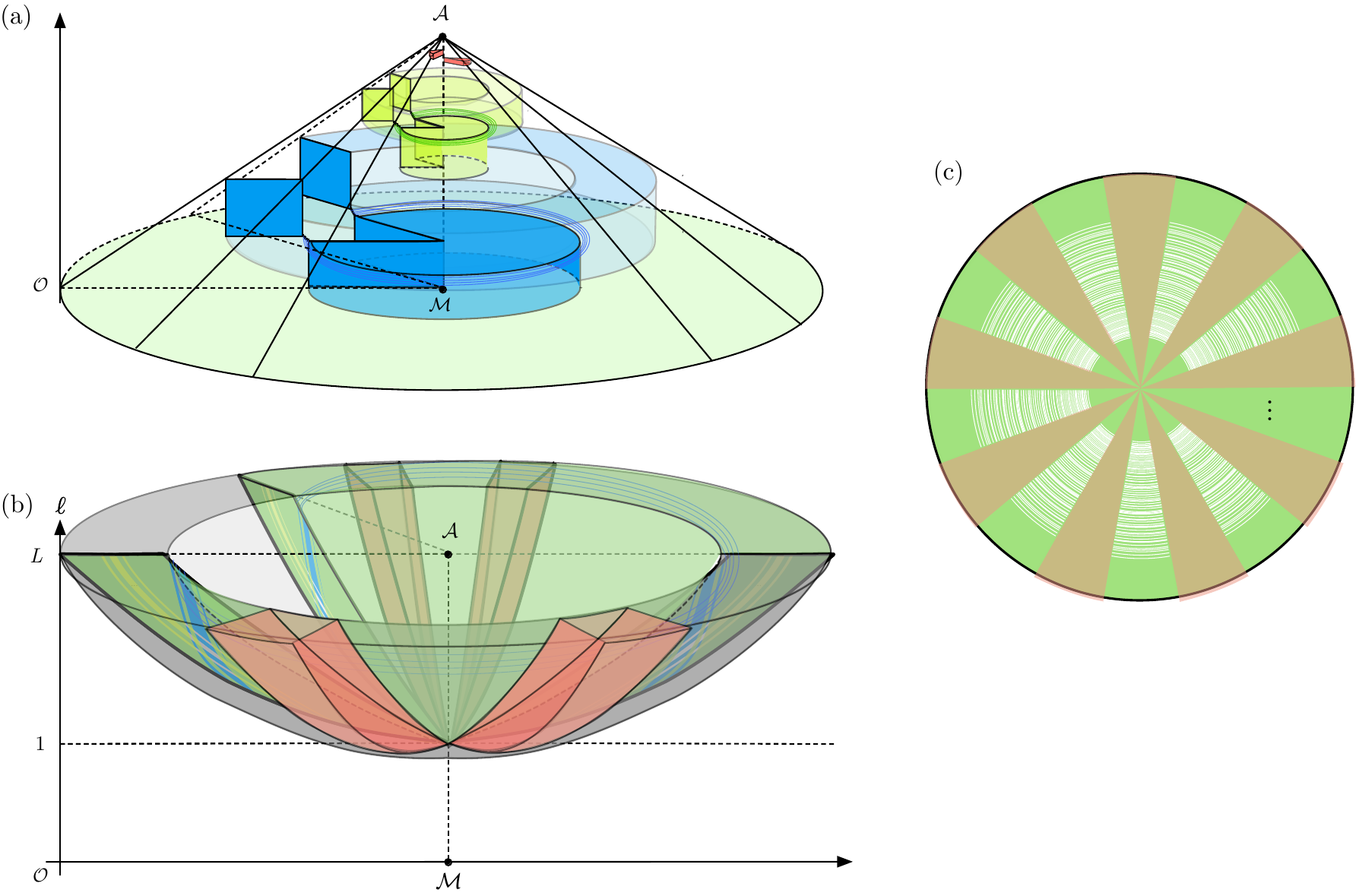}
    \caption{(a) The 3-dimensional family of obstacles described in Theorem \ref{th:3d_example}. (b) The visibility region of the family in Part (a). (c) The projection of the visibility region (clear region) in Part (a) to a horizontal plane.}
    \label{fig:3d_example}
\end{figure}

\begin{theorem}\label{th:3d_example}
    There is a family $\mathcal{B}_n$ of $n$ obstacles in the space, each an ACW-box, whose visibility region $\mathcal{V}_L\setminus\mathcal{B}_n^-$ has $\Theta(n^3)$ connected components. Moreover, the catenary visibility domain of $\mathcal{B}_n$, which corresponds to the projection of $\mathcal{V}_L\setminus\mathcal{B}_n^-$ onto the $(g,\theta)$-plane, consists of $\Theta(n^3)$ connected components.  
\end{theorem}
\begin{proof}
    Each ACW-box in $\Gamma$ can be described by $R_{[\theta_i,\theta_f]}=\bigcup_{\theta_i\leq \theta \leq ,\theta_f}R_\theta$, where $R_\theta$ is an orthogonal rectangle in $\Delta_\theta$ and all rectangles $R_\theta$ coincide if all triangles $\Delta_\theta$ with $\theta \in [\theta_i,\theta_f]$ were superimposed. 
    
    The family of obstacles is defined using the family of orthogonal rectangles $\mathcal{R}_{n/2}$ in Theorem \ref{th:2d_example} as follows: Let $\theta_i=2\pi i/n$ for $0\leq i < n$ and $\{R(i):1\leq i \leq n-1 \text{ odd}\}$ is a family of disjoint orthogonal rectangles in $\Delta$, each with upper left vertex in $OA$ and lower right vertex in $AM$ close enough to $A$ so that they do not intersect any rectangles in $\mathcal{R}_{n/2}$ (see Figure \ref{fig:3d_example}). Note that the visibility region ${R(i)}^-$  of $R(i)$ is empty. Then
    \begin{equation*}
        \mathcal{B}_n=\{R_{[0,\theta_{n-1}]}:R\in \mathcal{R}_{n/2}\}\cup\{R(i)_{[\theta_i,\theta_{i+1}]}:1\leq i \leq n-1 \text{ odd} \}.
    \end{equation*}
    
    Then the projection of the visibility region $\mathcal{B}^-_n$ onto $\mathcal{D}_L$ is formed by $\Theta (n^3)$ annular sectors as follows:
    \begin{equation*}
        \mathcal{B}^-_n=\{I_{[\theta_i\theta_{i+1}]}:1\leq i \leq n-1 \text{ even} ,I\in\mathcal{I}\},
    \end{equation*}
    where $\mathcal{I}$ is the set of $\Theta (n^2)$ closed intervals in the projection of $\mathcal{R}_{n/2}$ to the $g$-axis. 
\end{proof}

Although the family of obstacles $\mathcal{B}_n$ has the property that it actually yields only a constant number of distinct instances of the 2D-CV problem, it is possible to modify it to produce $O(n)$ distinct instances: we can carefully start the \emph{large} obstacles at distinct values of $\theta$, instead of all at the same value of $\theta$. 

\section{Conclusions and future work}

The integration of real-time geometric computation and robotic path planning represents a fundamental challenge in advanced marsupial autonomy. In current algorithms, the need for time-consuming queries regarding
the existence of collision-free catenaries is a major bottleneck \cite{martinez2023path}.     In this paper, we reduce the problem of computing obstacle-free catenaries to reporting horizontal and vertical intervals in arrangements of pseudosegments. This geometric framework enables us to efficiently answer queries in path planning tasks for robotic systems with tethers modeled by catenary curves.

In what follows, we discuss several improvements and generalizations of our model, some of which open up directions for future research.

\paragraph{Topological sweep.}

The standard topological search algorithm \cite{edelsbrunner1986topologically} can handle pseudolinear arrangements \cite{felsner2017pseudoline}. If applied here, it would replace the $O((n+k)\log n)$ runtime of the algorithm in section \ref{subsect:computing_visibility_range} with $=(n^2+k)$, which is asymptotically smaller when $k>\frac{n^2}{\log n}$, albeit with significant technical caveats. To handle pseudo-segments, the layout is treated as a partial layout of pseudo-lines: each pseudo-segment is conceptually extended to an infinite pseudo-line, the scan is performed on the resulting layout, and only the characteristics of the original parts are reported. Unlike infinite lines, segments introduce event points that are not intersections, specifically, the start and end points of the segments, and the triangles used by the scan may be incomplete, thus the approach must be adapted. Topological scanning avoids event sorting and handles all intersections in $O(n^2+k)$ time, which is asymptotically optimal. Its key requirements are that the curves are $x$-monotonic, that any two curves intersect at most once, and that the arrangement is in general position.
In our case, these combinatorial conditions are satisfied, and the ‘cut-and-horizon’ data structure used in the paradigm is based exclusively on them, rather than on straightness. However, topological sweeping is already a delicate process and, in the case of curved pseudo-segments, maintaining the cut-off line requires careful handling of comparisons between curved edges. Therefore, although in theory it can be applied without difficulty to pseudo-segments, it is used less frequently than planar sweeping for curved inputs, 
since robust comparison along the cut is more delicate in practice.

\begin{figure}[htb]
    \centering
    \includegraphics[width=.8\linewidth]{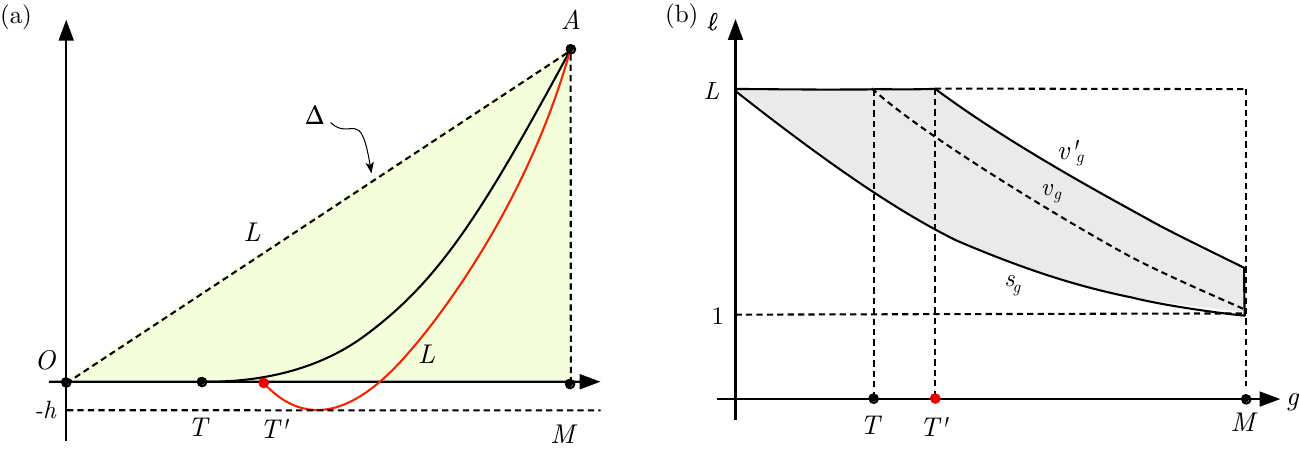}
    \caption{(a) The catenary can reach the line $y=-h$. (b) The new visibility region $V_L$.}\label{fig:newsag}
\end{figure}

\paragraph{Tether beneath the UGV.} In our study, we model the UGV as a point and assume the tether does not contact the ground in a point other than its launch point, where the ground is defined as the line $y=0$. Therefore, the vertex of the catenary is always located at or to the left of the launch point $G=(g,0)$. More concretely, this depends on the position of the transition point $T=(t,0)$ that separates the launch points where the tether can be extended to maximum length $L$, from the launch points where the maximum length (to which the tether can be extended) is bounded by the moment when the tether  starts laying on or going below the ground. Catenaries connecting points $G$ and $A$ have their vertices to the left of $G$ when $g\leq t$ or exactly on point $G$ when $t<g<m$, being $M=(m,0)$. Now consider a more realistic model in which the tether is anchored someplace on the UGV at distance $h$ from the ground. This situation allows for a sag in the tether, where the vertex may be located to the right of the launch point $G$. For simplicity, we keep the launch points on the $x$-axis and bound the maximum sag of the catenary by the line $y = -h$. The transition point is now denoted as $T'=(t',0)$ located to the right of $T$ (see Figure \ref{fig:newsag}(a)). For catenaries connecting $G$ and $A$, the location of the vertex depends on $g$: if $g < t$, the vertex is to the left of $G$; if $t \leq g \leq t'$, the vertex can be any point $(v_1, v_2)$ with $-h \leq v_2 \leq 0$ and $t\leq v_1\leq t'$; and if $t' < g < m$, the vertex lies on the line $y = -h$.
The only difference with the initial scenario is that the new visibility region contains the previous one with a different upper boundary $v'_g$ positioned higher than $v_g$ (see Figure \ref{fig:newsag}(b)).  The curve $v'_g$, now defined as the function representing the length of a catenary connecting $G$ and $A$ whose vertex lies on $y = -h$, extends the visibility region $V_L$ above. Our results still hold under this new modeling as long as there are no obstacles between the precise launching line and the ground. This makes sense in a real scenario as otherwise such obstacles would obstruct the UGV's movement.

\begin{figure}[htb]
    \centering
    \includegraphics[width=0.9\linewidth]{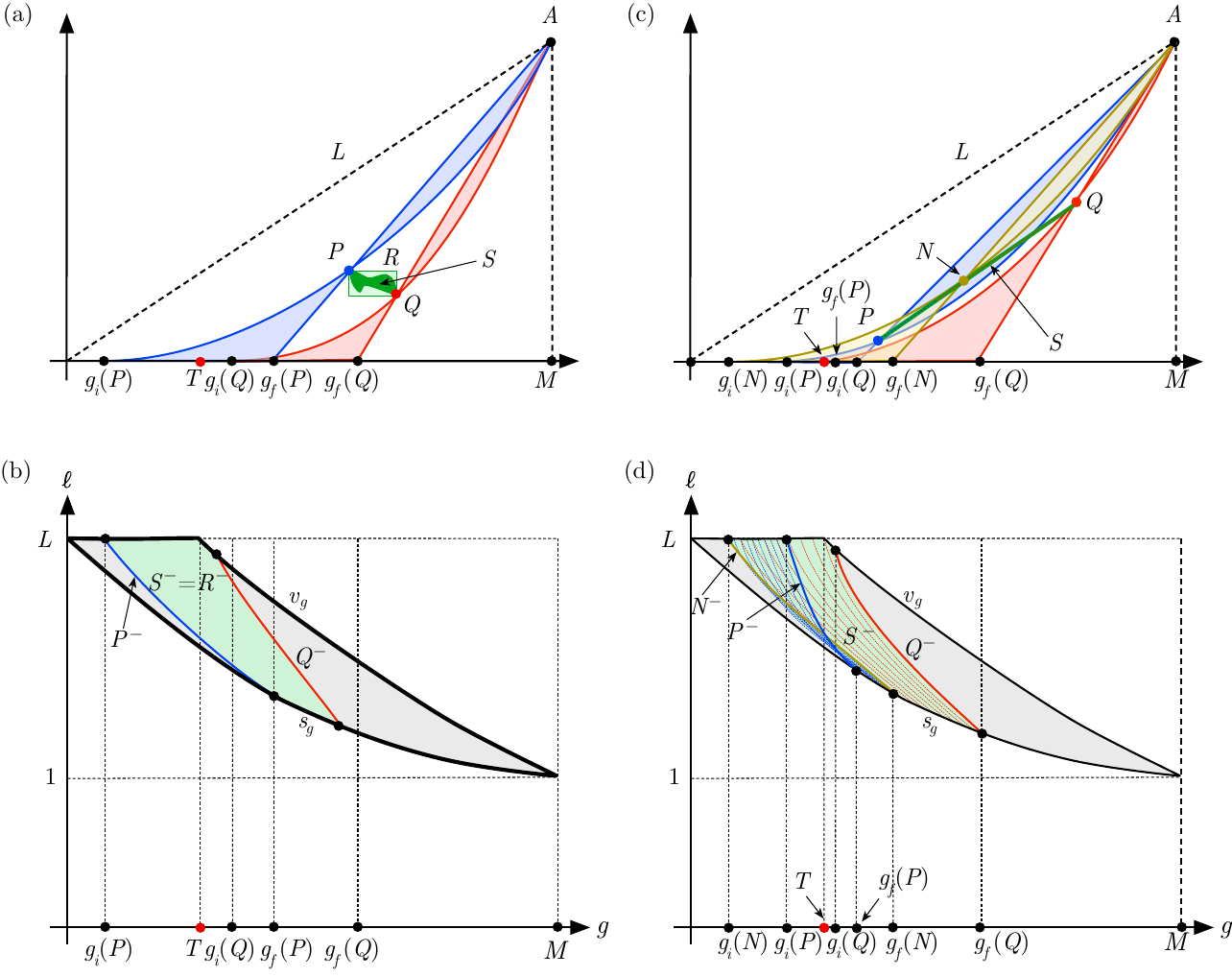}
    \caption{(a-b) The region obstructed by an obstacle $S$ (which could be the segment $PQ$), that coincides with the region obstructed by the orthogonal rectangle $R$ containing $S$. (c-d) The region obstructed by a segment $PQ$ with positive slope when the ``first obstruction'' point (from left to right) $N$ is an interior point of  $PQ$.}
    \label{fig:other_obstacles}
\end{figure}

 \paragraph{Handling generic obstacles.} The key observation that makes our strategy work is that, for orthogonal rectangular obstacles, the problem reduces to working with an arrangement of pseudosegments. It is possible to extend our approach to other families of obstacles, for example nonincreasing curves, by noting that the proof of Theorem \ref{th:one_rectangle} actually asserts the following statement. 
\begin{theorem}\label{th:one_decrcurve}
  Let $PQ\subset\Delta$ be a segment with nonpositive slope, $R$ the orthogonal rectangle with diagonal $PQ$, and $C$ a continuous curve from  $P$ to $Q$ contained $R$. Then $C^-=R^-$, the set of all points in $V_L$ between $P^-$ and $Q^-$.  Moreover, if $C$ is nonincreasing, then $C^-$ is the disjoint union of the curves $U^-$ with $U\in C$.
\end{theorem}

Note that this includes the case when the obstacle is the segment $PQ$. A direct implication of Theorems \ref{th:one_rectangle} and \ref{th:one_decrcurve} provides the most general setting for which our approach still applies  (see Figure \ref{fig:other_obstacles}a-b) as stated in the following corollary.
\begin{corollary}\label{th:max_obstacle}
 
  Let $PQ\subset\Delta$ be a segment with nonpositive slope, $R$ the orthogonal rectangle with diagonal $PQ$, and $S$ any connected obstacle such that $P,Q\in S\subseteq R$. Then $S^-=R^-$.
\end{corollary}
This result extends our solution to the 2D-CV problem to families of obstacles satisfying the conditions of Theorem \ref{th:max_obstacle}, which in particular includes families of segments with nonpositive slopes. In contrast, the region obstructed by a segment with positive slope is significantly more complicated. When the segment $S=PQ$ has positive slope, the first point $(g,\ell)\in V_L$ in lexicographic order such that the catenary $c(g,\ell)$ intersects the segment $PQ$ could be tangent to $PQ$ at an interior point $N$, see Figure \ref{fig:other_obstacles}c. In this case, the pseudosegments $U^-$ where $U\in PN$ intersect in pairs forming an envelope that constitutes the left boundary of the region $S^-$, see Figure \ref{fig:other_obstacles}d. A theoretical study of such boundary's complexity along with an understanding of how the boundaries of the regions obstructed by two or more disjoint segments with positive slope may intersect would be required to decide whether or not our approach or a similar strategy may be adapted to solve the 2D-CV problem in this case. We leave this investigation as a direction for future work, emphasizing that in practical scenarios, any obstacle can be approximated using orthogonal rectangles, to which our procedure readily applies.

\newpage
\bibliographystyle{plain}
\bibliography{references}

\newpage

\section*{Appendix. Notation}
\begin{table}[hbt!]
    \centering
    \begin{tabular}{ll}
        UGV, UAV & Unmanned Ground Vehicle, Unmanned Aerial Vehicle \\
        2D-CV, 3D-CV & 2- and 3-Dimensional Catenary Visibility (problem), respectively\\
        ACW & Annular Cylindrical Wedge\\
        $\cal L$ & UGV's motion, typically the $x$-axis\\ 
        $L>0$ & maximum tether extension (maximum length of a catenary)\\ 
        $A=(\sqrt{L^2-1},1)$ & UAV's target point (2D)\\
        $\mathcal{A}=(\sqrt{L^2-1},0,1)$ & UAV's target point (3D)\\
        $\pi$ & vertical plane containing the obstacles (2D)\\
        $\cal R$ & set of orthogonal rectangular obstacles (obstacles in 2D)\\
        $\cal B$ & set of ACW-boxes (obstacles in 3D)\\
        $R,R_1, R_2, \dots$ & orthogonal rectangular obstacles\\
        $n$ & number of obstacles\\
        $\cal C$ & a collision-free curve\\
        $G=(g,0)$ & a position of the UGV, $0\leq g \leq \sqrt{L^2-1}$\\
        $O=(0,0)$ & furthest position of the UGV from which the UAV may be reachable (to the left of $A$)\\
        $O'=(2\sqrt{L^2-1},0)$ & furthest position of the UGV from which the UAV may be reachable (to the right of $A$)\\
        $M=(\sqrt{L^2-1},0)$ & midpoint of $OO'$ and point on $\cal L$ directly below $A$\\
        $\mathcal{M}=(\sqrt{L^2-1},0,0) $ & projection of $\mathcal{A}$ onto the $xy$-plane\\
        $\ell$ & length of a catenary \\
        $c(g,\ell)$ & the catenary passing through $G=(g,0)$ and $A$ with length $\ell$ from $G$ to $A$\\
        $(g,\ell)$ & point of the visibility plane corresponding to the catenary $c(g,\ell)$\\
        $s_g$ & length of the segment $GA$ \\
        $v_g$ & length of the catenary through $A$ with vertex $G=(g,0)$ from $G$ to $A$\\
        $T=(t,0)$ & transition point, i.e., unique point on $\cal L$ with $v_t=L$\\
        $\Delta$ & $\triangle OAM$, only obstacles intersecting this solid triangle obstruct the tether\\
        $V_L$ & visibility region in the visibility plane when no obstacles are present in the obstable plane.\\
        $\beta_1,\beta_2,\beta_3$ & the three pseudosegments forming the boundary of $V_L$\\
        $P,Q,Z$ & points in $\Delta$\\
        $P^-,R^-,\cal R^-$ & the region obstructed by the point $P$, the orthogonal rectangle $R$, the set of obstacles $\cal R$\\
        $F_g$ & $\{\ell: c(g,\ell) \text{ is collision-free}\}$, set of feasible lengths for $g$ \\
        $g$ has full visibility & $F_g=[s_g,\min(L,v_g)]$ \\
        $g$ has null visibility & $F_g=\emptyset$ \\
        $g$ has  partial visibility & $F_g\neq \emptyset$ and $F_g\neq [s_g,\min(L,v_g)]$ \\
        shadow of $\cal R$ & $\{g\in [0,\sqrt{L^2-1}]:F_g=\emptyset\}$\\
        shadow of $\cal B$ & $\{(g,0)_\theta\in \mathcal{D}_L:F_g=\emptyset \text{ in } \Delta_\theta\}$\\
        visibility range of $\cal R$ & $[0,\sqrt{L^2-1}]$ minus the shadow of $\cal R$ \\ 
        visibility domain of $\cal B$ & $\mathcal{D}_L$ minus the shadow of $\cal B$ \\ 
        $\mathcal{R}_n$ & a special set of $n$ rectangular obstacles whose visibility range has $\Theta(n^2)$ components\\
        $\mathcal{B}_n$ & a special set of $n$ ACW-boxes whose visibility domain has $\Theta(n^3)$ components\\
        $\mathcal{C}_L$ & circumference in the $xy$-plane with center $\mathcal{M}$ passing through $\mathcal{O}$\\
        $\mathcal{D}_L$ & disk with boundary $\mathcal{C}_L$\\
        $\mathcal{O}=O_0=(0,0,0)$ & furthest position of the UGV from which the UAV may be reachable in the 3D model\\
        $O_\theta$ & point on $\mathcal{C}_L$ such that $\angle \mathcal{OM}O_\theta=\theta$\\
        $\Delta_\theta$ & $\triangle \mathcal{O_\theta MA}$, identified with $\Delta OMA$ for each $\theta$\\
        $\Gamma_L$ & vertical cone with base $\mathcal{D}_L$ and vertex $\mathcal{A}$\\
        $(x,y)_\theta, G_\theta,V_{L,\theta},c(g,\ell)_\theta$ & objects in $\Delta_\theta$ corresponding to $(x,y), G,V_L,c(g,\ell)$ in $\Delta$\\
        $s_g(\theta),v_g(\theta)$ & lengths in $\Delta_\theta$ corresponding to $s_g,v_g$ in $\Delta$\\
        $\mathcal{V}_L$ & $\bigcup _{\theta\in[0,2\pi)}V_{L,\theta}$, the visibility region when no obstacles are present (3D)\\
        $R_{[\theta_i,\theta_f]}\in\Gamma_L$ & ACW-box whose vertical sections are the orthogonal rectangle $R\in\Delta_\theta$ for $\theta_i\leq \theta \leq\theta_f$\\
        $R^-_{[\theta_i,\theta_f]}\in\mathcal{V}_L$ & region obstructed by $R_{[\theta_i,\theta_f]}$\\
    \end{tabular}
\end{table}

\end{document}